# A new model for the effective thermal conductivity of polycrystalline solids

Fergany Badry[1], Karim Ahmed[1,2]
[1] Nuclear Engineering Department, Texas A&M University, College Station, TX, USA, 77843.
[2] E-mail: karim.ahmed@tamu.edu

**Abstract**

We introduce a novel model for the effective thermal conductivity of polycrystalline solids based on the thin-interface description of grain boundaries (GBs). In contrast to existing models, our new model treats a GB as an autonomous "phase" with its own thermal conductivity. The Kapitza resistance/conductance of a thin interface is then derived in terms of the interface thermal conductivity and width. In turn, the effective thermal conductivity of polycrystals is derived in terms of grain size, grain and GB conductivities, and GB width. This treatment allows the model to simulate the change of Kapitza resistance/conductance with segregation/doping, GB structure/phase transition, or GB decohesion. Moreover, since the model assumes a finite width for GBs, it is expected to give better predictions than its sharp-interface-based counterparts for nanoscale grains. The predictions of the new model deviate from the corresponding ones from existing models by 1-100% as the grain size approaches the GB width. High-fidelity finite-element simulations were conducted to validate the predictions of the new model. These simulations proved the higher accuracy of the new model. We also discuss how to generalize this treatment to other types of interfaces in heterogeneous materials. The advantages and limitations of the new model are summarized, and some future directions are highlighted.

## 1. Introduction

Thermal conductivity is one of the most important physical properties of materials. It is the main attribute that determines which material can be used in several industrial and technological applications. For instance, materials for thermal barrier coatings and thermoelectric materials must have low thermal conductivity, while materials for electronic devices and nuclear fuels must possess high thermal conductivity.

As most physical properties of solids, thermal conductivity is sensitive to the underlying microstructure[1-11]. The main reason for this dependency is the effect of interfaces in heterogeneous materials on the overall heat transfer. Interfaces usually act as obstacles for heat conduction by scattering heat carriers. For example, polycrystalline solids generally show lower thermal conductivity than their corresponding bulk single crystals. The reduction increases with decreasing grain size. Nonetheless, interfaces can also be engineered to improve the overall conductivity of heterogeneous materials by precipitating out a highly-conducting phase, adding highly-conducting intermediate layers, or doping[12, 13].

Understanding the interaction between heat carriers and interfaces is a very active area of research[8-37]. A comprehensive understanding of this process still eludes the community[32, 35, 37]. For most purposes, however, it is the effective/overall thermal conductivity of the inhomogeneous material that is of interest. Nonetheless, accurate predictions of the effective thermal conductivity require a precise description of the interfacial heat transport.

Our main focus here is on the effect of GBs on the effective thermal conductivity of polycrystalline solids. Nevertheless, the main treatment and concepts can be adapted to other types of interfaces and heterogeneous solids. Several models and simulation techniques were developed for predicting the conductivity of polycrystals [8-10, 12-37]. These models can be classified into two main categories, e.g. continuum and atomistic/particulate models. The acoustic mismatch model, diffuse mismatch model[38], and molecular dynamics simulations preserve the underlying particulate nature of heat carriers [12, 13, 32, 35, 37]. Continuum models and finite-element simulations, on the other hand, utilize specific constitutive laws and interfacial boundary conditions that coarse grain the atomistic details [8-11, 15-22, 39, 40].

Continuum models have the advantages of simplicity of implementation and usage, ease of interpretation of results, and capability of verification and validation using relatively simple experiments. In these models, the most crucial parameter that accounts for interfacial transport is the thermal boundary (Kapitza) resistance/conductance. This parameter, however, can only be determined from atomistic models or advanced experiments. Several studies have shown that its value depends on GB energy, GB misorientation angle, GB excess volume, and strain energy[30-35]. Moreover, as for the case of heterojunctions in semiconductors [12, 13], one expects segregation and interface structure/phase transition to affect the value of the Kapitza resistance as well. Yet, all these factors are absent from current analytical and continuum models[8-10] commonly employed in literature[10, 14-29]. Moreover, since these models are based on the sharp-interface description of GBs, their predictions for nano-sized polycrystalline solids are questionable.

Here, we introduce a new continuum model that alleviates the shortcomings of existing models mentioned above. It is based on the thin-interface description of GBs/interfaces. This description is common in the thermodynamic theory of heterogeneous materials, which treats GBs and interfaces as autonomous "phases"[41]. In deriving this model, we introduce an expression for the effective/average Kapitza resistance/conductance of a thin interface. This new expression allows continuum models to account for the change of thermal resistance/conductance of an interface due to segregation/doping, localized phase transition, or confined mechanical damage. The predictions of the new model were validated using high-fidelity finite-element simulations. These simulations demonstrated that the new model gives more accurate results than its sharp-interface-based counterparts, particularly for nano-sized grains.

This paper is organized as follows. In section 2, we introduce our new continuum treatment of interfacial transport, and our derivation of the effective thermal conductivity of a polycrystalline material. In that section, we also discuss our numerical approach that is then utilized to validate our model. In section 3, we present and discuss the main results. Lastly, our concluding remarks and future directions are summarized in section 4.

## 2. Continuum Modeling of the Effect of GBs on Heat Conduction in Polycrystalline Solids

### 2.1 A new model for the effective thermal conductivity of polycrystalline solids

Before we introduce our new model that is based on the thin interface description of grain boundaries, we briefly review first the classical models that employ sharp-interface description. By using the averaging grain theory [42], Nan and Birringer proposed in 1998 an analytical model to show the effect of grain size on the effective thermal conductivity of polycrystals. By incorporating the concept of the Kapitza resistance into an effective medium approach, this model describes the effective thermal conductivity for common polycrystals with isotropic, equisized spherical crystallites [1, 9].

$$k = \frac{k_b}{1 + \frac{2k_b R_k}{d}} \tag{1}$$

Where, $k_b$ is the single crystal thermal conductivity, $R_k$ is the Kapitza thermal resistance from grain boundaries, and $d$ is grain size.

This model has been revised by Yang et al. in 2002 by noting that each grain boundary region is shared by two grains[1, 8]. Therefore, instead of Eq. (1), the effective thermal conductivity is given by [8]

$$k = \frac{(k_b)}{1 + \frac{k_b R_k}{d}} \tag{2}$$

More recently, Palla and Giordano [11] generalized the above models to anisotropic materials and considered both the highly- and lowly-conducting interfaces. While this generalization added a significant contribution to the earlier work, the dependence of the effective thermal conductivity on grain size remained the same.

The classical models summarized above are based on a sharp-interface description of the interface. In such description, it is assumed that, while the heat flux is continuous, there is a discontinuity in the temperature field across the grain boundary. Nonetheless, as we mentioned before, at small scale, the thin-interface description provides more natural and accurate treatment of the interfacial region. Moreover, it accounts in a straightforward manner for the possibility of drastic physical property change in this region due to doping, segregation, and/or interface phase transition. Therefore, we develop here the thin-interface model of heat transfer across grain boundaries in solids. The descriptions of the sharp- and thin- interface treatments are schematically illustrated in Fig. 1. In contrast to the sharp-interface description, the thin-interface description treats the interfacial region as autonomous with its own thermodynamic properties (e.g. a surface phase). In its simplest form, which will be considered here, it implies the continuity of both heat flux and temperature across the boundaries; effectively transforming an imperfect interface into two perfect interfaces separated by a distance equal to the interface width (as shown in Fig. 1).

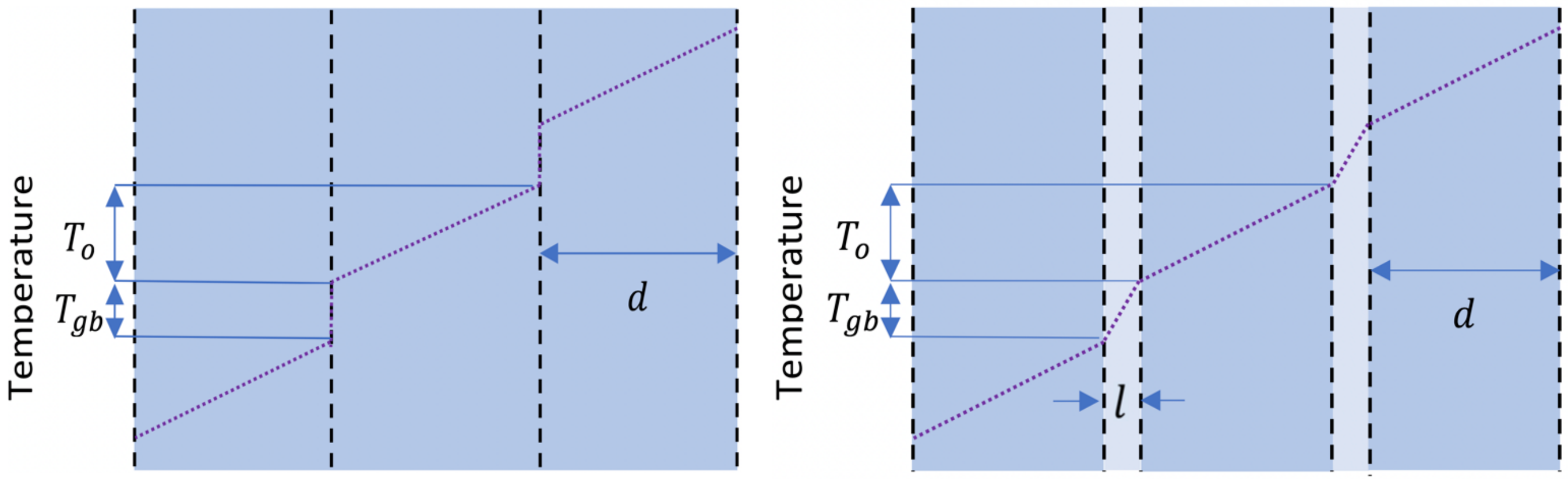

Figure 1. A schematic representation of the sharp (left) and thin-interface (right) descriptions of temperature drop across grain boundaries in solids. Dashed lines represent grain boundaries.

In order to derive an equation for the effective thermal conductivity, we generalize here the procedure carried out by Yang et al [8]for the sharp-interface case to the thin-interface case. The average temperature drop across a grain is $T_o$. $T_{gb}$ is the temperature drop across a GB because of its thermal (Kapitza) resistance. Similar to the procedure in [1], [8], we examine the temperature profile across a polycrystalline system in response to an applied heat flux q. The heat flux is assumed to follow the classical Fourier's law,

$$q = -k\frac{dT}{dx} \quad (3)$$

As evident from Fig. 1, the total temperature drop across the grain and its boundary is

$$T_{tot} = T_o + T_{gb} \quad (4)$$

By substituting Eq. (4) into Eq. (3) and rearranging, we obtain

$$k = \frac{-q(d + l)}{T_o + T_{gb}} \quad (5)$$

Where $\frac{T_o + T_{gb}}{(d+l)}$ is equivalent to $\frac{dT}{dx}$.

Similarly, by applying Fourier's law inside the grain, one has

$$T_o = \frac{-qd}{k_b} \tag{6}$$

By definition, the Kapitza resistance is the ratio between the temperature drop across the interface to the heat flux passing through it[43], e.g.,

$$R_k = -\frac{T_{gb}}{q} \tag{7}$$

Hence,

$$T_{gb} = -qR_k \tag{8}$$

By combining Eqs. (5), (6), and (8), we arrive at

$$k = \frac{-q(d+l)}{\frac{-qd}{k_b} - qR_k} \tag{9}$$

Or equivalently,

$$k = \frac{(d+l)}{\frac{d}{k_b} + R_k} \tag{10}$$

As expected, in the limit where the interface width vanishes, Eq. (10) reduces to Eq. (2) and we recover Yang et. al. model. It is also worth noting that one can arrive at an equation similar to Eq. (10) based on the harmonic-average formula of layered composites [44]. This can be achieved by treating both the grains and grain boundaries as "phases" with their own thermal conductivities, as shown in Fig. 2.

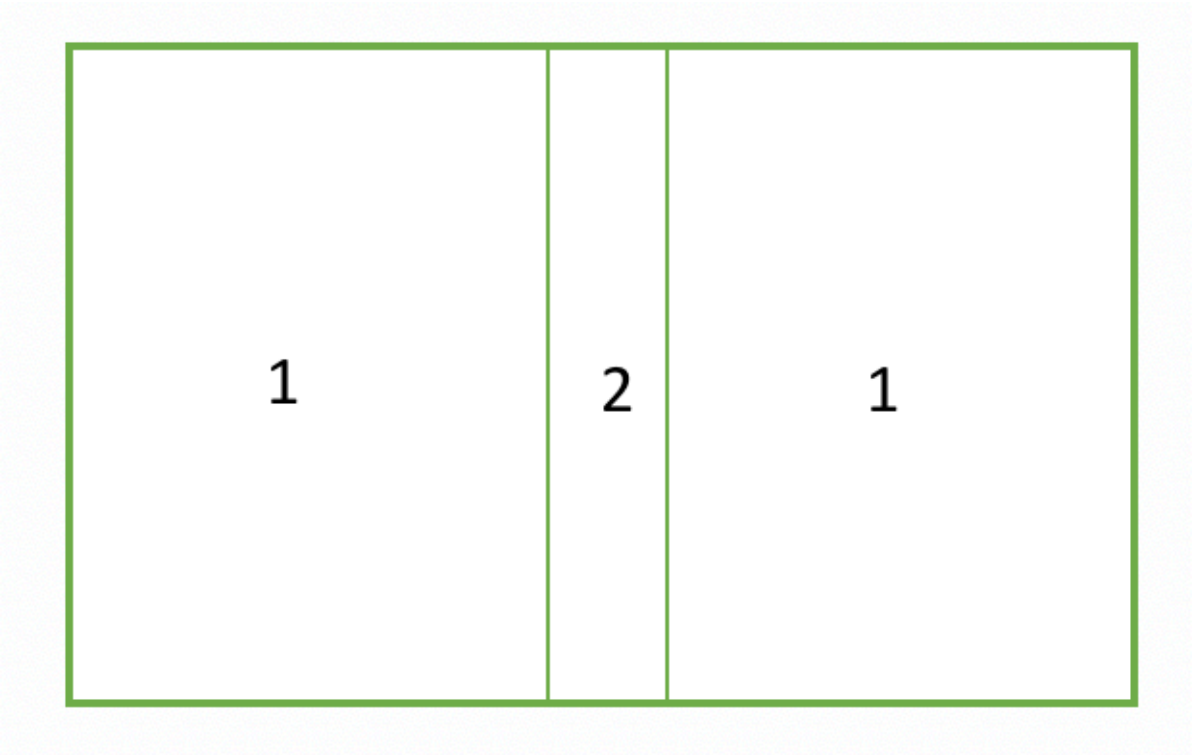

Figure 2. A schematic representation of a bicrystal as a layered composite of two phases, e.g., the grain "phase" (1) and the grain boundary "phase" (2)

In such case, the average thermal conductivity is simply given by,

$$k = (\frac{\emptyset_1}{k_1} + \frac{\emptyset_2}{k_2})^{-1} \tag{11}$$

Where, $\emptyset_1$, and $\emptyset_2$ represent bulk volume fraction, and grain boundary volume fraction, respectively; i.e., $k_1 = k_{\mathrm{b}}$, and $k_2 = k_{\mathrm{gb}}$.

The first phase (grain) volume fraction is given by

$$\emptyset_1 = d/(d + l) \tag{12}$$

The second phase (grain boundary) volume fraction is

$$\emptyset_2 = l/(d + l) \tag{13}$$

By substitution of Eqs. (12) and (13) in Eq. (11), we have

$$k = \frac{1}{\frac{d/(d+l)}{k_b} + \frac{l/(d+l)}{k_{gb}}} \tag{14}$$

Or equivalently,

$$k = \frac{(d + l)}{\frac{d}{k_b} + \frac{l}{k_{gb}}} \tag{15}$$

The similarity between equations (10) and (15) is obvious. By comparing Eqs. 10 and 15, one can relate the interface/Kapitza resistance of the sharp-interface to the thermal conductivity of the thin interface as

$$R_k = \frac{l}{k_{gb}} \tag{16}$$

Such expression, however, is valid only for the case of constant thermal conductivity of the grains and their boundaries. The generalization of this expression to variable GB conductivity will be considered in the next subsection.

So far, we have shown using two different approaches how to derive the effective thermal conductivity of a polycrystalline solid based on the thin-interface analysis. In such description, the effective thermal conductivity was shown to be

$$k = \frac{(d + l)}{\frac{d}{k_b} + R_k} \tag{17}$$

In contrast to the classical sharp-interface-based models of the effective conductivity, here there is a dependence on both the grain size and the grain boundary width. It is interesting to note that the thin-interface-based model always predicts a higher value of the effective thermal conductivity than its sharp-interface-based counterpart. By comparing the new model prediction (Eq. 17) with Yang et al model prediction (Eq. 2), one can quantify the difference as

$$D_f = \frac{l}{d} \tag{18}$$

where $D_f$ is the fractional difference between both analytical models, e.g., $\frac{k_{new\ model} - k_{Yang\ et\ al\ model}}{k_{Yang\ et\ al\ model}}$. As one might expect, the difference increases as the ratio between the grain boundary thickness and grain size increases, with a maximum fractional difference of 100% when the ratio equals one. The predictions of this new model will be validated using finite-element simulations in the next section.

The above thin-interface-based analysis can also be applied to a hetero-interface (interphase interface) between two phases with different bulk conductivities. For such system, in addition to the interfacial parameters, the conductivity, fraction, size and morphology of each phase will affect the effective value. Such generalization is discussed in Appendix A.

### 2.2 General formula for the effective Kapitza resistance of a thin interface

As we have shown above, an expression for the effective conductivity of a polycrystalline solid can be derived based on sharp- or thin- interface description of the grain boundary. To shed light into the connection between these two different treatments of interfaces, one must relate the Kapitza/interface thermal resistance of a sharp-interface to the thin-interface thermal conductivity and width. We derive here such a general formula for the effective thermal resistance of a thin interface.

For the case of steady-state heat conduction, and taking $x$ to be the direction normal to the grain boundary, on has, from Fourier's law,

$$q = -k(x)\frac{dT}{dx} = const \tag{19}$$

$$dT = -\frac{q}{k(x)}dx \tag{20}$$

For the sake of generality, we assume the thermal conductivity to be spatially dependent. By integrating this equation across the grain boundary/interface, e.g.,

$$\int_0^{T_{gb}} dT = \int_0^l \frac{q}{k(x)_{gb}}dx, \tag{21}$$

where $l$ , and $T_{gb}$ are the grain boundary width, and the temperature drop through the grain boundary, respectively. For steady-state conditions, we then have

$$\frac{T_{gb}}{q} = \int_0^l \frac{dx}{k(x)_{gb}}, \tag{22}$$

Based on the definition of the Kapitza resistance (recall Eq. (7)), one arrives at

$$R_k = \int_0^l \frac{dx}{k(x)_{gb}} \tag{23}$$

Thus, the Kapitza conductance is

$$G_k = \frac{1}{\int_0^l \frac{dx}{k(x)_{gb}}} \tag{24}$$

The validation of these formulas of the effective Kapitza resistance and conductance for lowly- and highly-conductive thin interfaces with variable conductivity will be verified using finite-element simulations in the next section.

### 2.3 Calculating the effective thermal conductivity of polycrystalline solids

In order to verify the expressions derived above for the thin interface resistance/conductance and the effective thermal conductivity, we utilize a computational approach that combine finite-element and phase-field methods [15, 18, 20, 22, 40, 45-50]. In such approach, constant temperatures $T_r$ & $T_l$ are applied on the right and left sides, respectively. The top and bottom boundaries (y-direction) are taken as adiabatic. Consequently, the thermal gradient points in the $x$ -direction. This configuration is schematically shown in Fig. 3. The heat flux, $q$, profile in the system is obtained by solving the steady-state conduction equation, e.g.,

$$\nabla.(k\,\nabla T) = 0 \tag{25}$$

where $k$ is the thermal conductivity, which varies spatially throughout the domain to account for the underlying microstructure. For the simple case of a bicrystal, which is considered first to validate the model, the location of the grain boundary is known a priori and Eq. (25) can be solved directly. However, for the case of complex grain structures in polycrystalline solids that may undergo grain growth, we use the phase-field method to represent the microstructure. Specifically, the phase-field model of grain growth [51-55] is utilized to distinguish between grains and grain boundaries, and to track the microstructure evolution during grain growth. In that model, phase-field variables $\eta_i$ are assigned such that their values indicate the type of region (e.g., gains, or GBs). The thermal conductivity is then assigned based on these values. Lastly, the overall effective thermal conductivity can be calculated from:

$$K_{eff} = \frac{q\,L}{T_r - T_l}, \tag{26}$$

where $q$ is the average heat flux, $T_r$ & $T_l$ are the temperatures at right and left boundaries respectively, and $L$ is the width of the simulation domain (see Fig. 3).

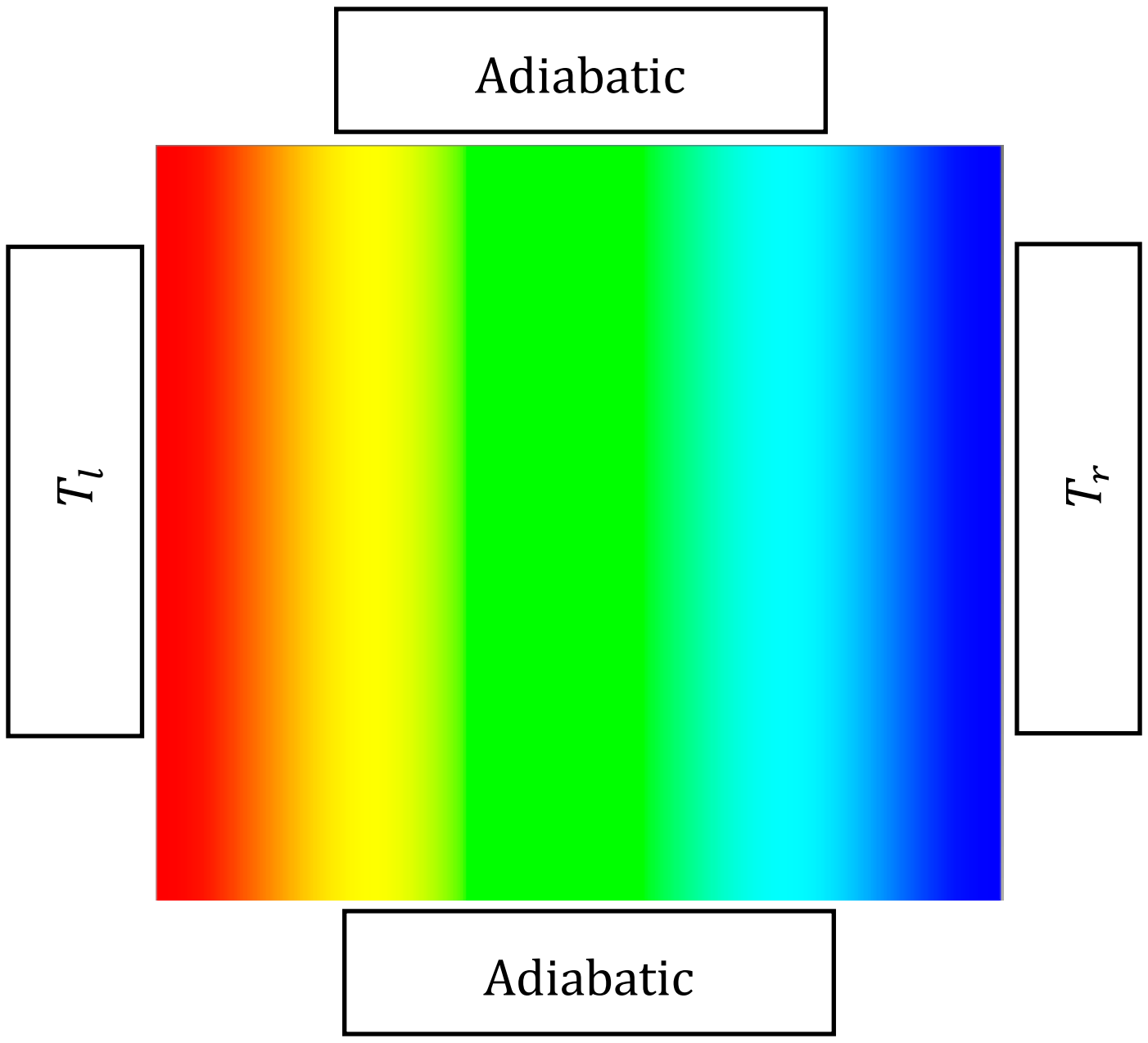

Figure. 3. A schematic illustration of the simulation domain used to calculate the effective thermal conductivity.

## 3. Results and discussion

### 3.1 Validation of the Kapitza resistance/conductance formula for a thin GB

While in most cases interfaces impede heat transport, the concepts of interface engineering can be utilized to optimize the thermal conductivity of the interface [56]. For instance, it was recently shown that mass-graded interlayers enhance the thermal transport through solid-solid interfaces[13]. Moreover, in some situations, the interface conductivity might be higher than the bulk conductivity due to doping and segregation of high thermal conductivity elements in multi-component systems, occurrence of an interface phase transition, or the presence of high density of point and/or line defects in the bulk phases because of irradiation or severe mechanical deformation. We demonstrate here that the derived expressions of Eqs. (23) and (24) are capable of describing accurately these two scenarios. Namely, Eq. (23) gives the Kapitza resistance of a lowly-conducting interface and Eq. (24) gives the Kapitza conductance of a highly-conducting interface. Furthermore, these relations are also valid even if the interface conductivity changes with position inside the interfacial region. In the coming subsections, we validate both analytically and numerically the above-mentioned relations for different spatial profiles of the interface thermal conductivity.

### 3.1.1 Kapitza resistance of the lowly-conducting GB

We consider three different interfacial profiles for the thermal conductivity of the GB in a bicrystal configuration as shown in Figure 4 below. For each case, we derive analytically the effective Kapitza resistance of the thin GB using Eq. (23). We then verify the results using finite-element simulations.

#### 3.1.1.1 Constant profile:

$$k(x)_{gb} = k_{gb} \qquad 0 \le x \le l \tag{27}$$

$k_{gb}$ is the constant GB thermal conductivity and its value was fixed at 0.1 of the bulk thermal conductivity value ($k_b$), which is taken to be 1 $W/m.K$ . By substituting Eq. (27) into Eq. (23), we obtain

$$R_k = \int_0^l \frac{dx}{k_{gb}} = \frac{l}{k_{gb}} \tag{28}$$

#### 3.1.1.2 Linear profile:

$$k(x) = \begin{cases} k_b(a + b\,x), & 0 \le x \le \frac{l}{2} \\ k_b(c + \mathrm{e}\,x), & \frac{l}{2} \le x \le l \end{cases} \tag{29}$$

Where $a, b, c, and\ e$ are constants. The interface thermal conductivity value reaches 0.1 of the bulk thermal conductivity at the middle of the GB. Again, by substituting Eq. (29) into Eq. (23) we obtain,

$$R_k = \int_0^{\frac{l}{2}} \frac{dx}{k_b(a + b\,x)} + \int_{\frac{l}{2}}^{l} \frac{dx}{k_b(c + \mathrm{e}\,x)} \tag{30}$$

Which upon integration gives,

$$R_k = \frac{l}{k_b}\left[\frac{1}{b}\ln\left(a + \frac{b\,l}{2}\right) - \frac{1}{b}\ln(a) + \frac{1}{\mathrm{e}}\ln(c + \mathrm{e}l) - \frac{1}{\mathrm{e}}\ln\left(c + \frac{\mathrm{e}\,l}{2}\right)\right] \tag{31}$$

The constants and Kapitza resistance value are listed in Table. 1

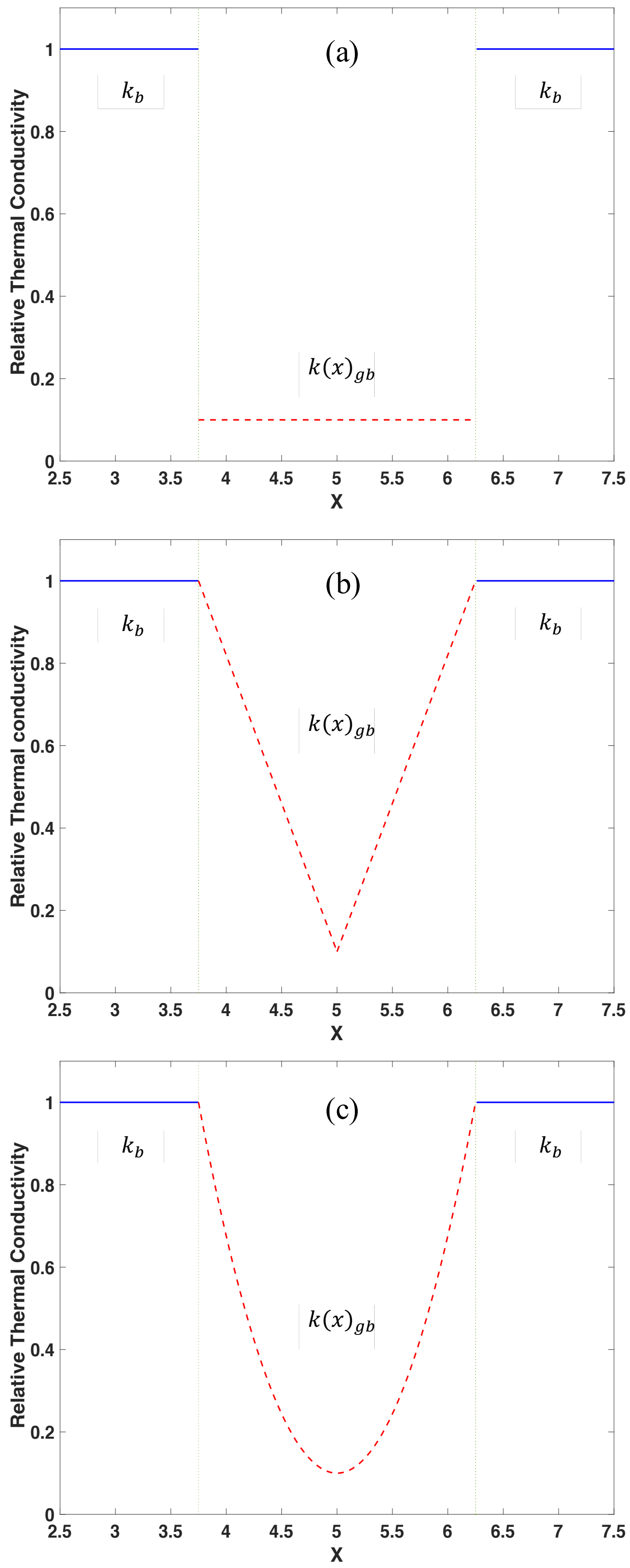

Figure 4. Different spatial profiles for the thermal conductivity of the lowly-conducting grain boundary: (a) constant, (b) linear, and (c) parabolic.

3.1.1.3 Parabolic profile

$$k(x) = k_b(a + bx^2) \qquad -\frac{l}{2} \le x \le \frac{l}{2} \tag{32}$$

Where $a\, and\, b$ are constants which are taking specific values to set the interface thermal conductivity as 0.1 of the bulk thermal conductivity at the middle of the GB. By substituting Eq. (32) into the new formula Eq. (23).

$$R_k = \int_{-\frac{l}{2}}^{\frac{l}{2}} \frac{dx}{k_b(a + bx^2)} \tag{33}$$

Direct integration yields,

$$R_k = \frac{2}{k_b\sqrt{ab}} \left[ \tan^{-1}\left(\frac{l}{2}\sqrt{\frac{b}{a}}\right) \right] \tag{34}$$

The constants and Kapitza resistance value are also listed in Table. 1

To verify the above results, we perform finite-element simulations of the bicrystal configurations shown in Fig. 4. In these simulations, the grain size was 10 $nm$ , the grain boundary width was 2.5 $nm$, and the bulk conductivity was 1 $W/m.K$. From the simulations, we calculate the Kapitza resistance directly from the flux and thermal gradient across the grain boundary as given by Eq. (7). The analytical and numerical values of the Kapiza resistance are included in Table 1 below.

Table 1: analytical and numerical Kapitza resistance results for lowly-conducting grain boundary where, $l = 2.5\ nm$ , and $k_b = 1\ W/m.K$

| Spatial dependence profile | Constants | | | | Kapitza resistance $(m^2.k)$/W | | |
|---|---|---|---|---|---|---|---|
| | | | | | New formula | FEM | |
| | a | b | c | e | | Δx = 0.5 nm | Δx = 0.025 nm |
| Constant | N/A | N/A | N/A | N/A | $2.5\times 10^{-8}$ | $2.5\times 10^{-8}$ | $2.5\times 10^{-8}$ |
| Linear | 1 | -1.8/$l$ | -0.8 | 1.8/$l$ | $6.396\times 10^{-9}$ | $6.3132\times 10^{-9}$ | $6.3960\times 10^{-9}$ |
| Parabolic | 0.1 | 3.6/$l^2$ | N/A | N/A | $1.0409\times 10^{-8}$ | $1.0362\times 10^{-8}$ | $1.0409\times 10^{-8}$ |

As evident from Table 1, the FEM numerical results agree well with the newly derived Kapitza resistance formula (Eq. (23)) and are in close agreement for all spatial profiles of the GB thermal conductivity. This agreement improves with a decrease in element size (or equivalently increasing the number of elements inside the GB). It is worthy to note that the spatial profile of the thermal conductivity of the GB affects the Kapitza resistance, with the constant profile leading to the highest value and the linear profile resulting in the lowest value.

### 3.1.2 Kapitza conductance of the highly-conducting GB

Again, we consider three different interfacial profiles for the thermal conductivity of the grain boundary in a bicrystal configuration as shown in Figure 5 below. For each case, we derive analytically the effective Kapitza conductance of the thin GB using Eq. (24). We then verify the results using finite-element simulations.

#### 3.1.2.1 Constant profile:

To represent the highly conductive interface, the GB thermal conductivity was fixed to be 10 times the bulk thermal conductivity value (see Figure 5-a). By combining Eqs. (24) and (27), we obtain

$$G_k = \frac{1}{\int_0^l \frac{dx}{k_{gb}}} \tag{35}$$

which immediately gives,

$$G_k = \frac{k_{gb}}{l} \tag{36}$$

#### 3.1.2.2 Linear profile:

With specific values of $a, b, c, and\ e$ constants (recall Eq. (29)), the GB thermal conductivity was fixed to be 10 times the bulk thermal conductivity value at the middle of the GB (see Fig. 5-b) . By combining Eqs. (24) and (29) we obtain

$$G_k = \frac{1}{\int_0^{\frac{l}{2}} \frac{dx}{k_b(a + b\,x)} + \int_{\frac{l}{2}}^{l} \frac{dx}{k_b(c + \mathrm{e}\,x)}} \tag{37}$$

By direct integration, we arrive at

$$G_k = \frac{k_b}{l\left[\frac{1}{b}\ln\left(a + \frac{b\,l}{2}\right) - \frac{1}{b}\ln(a) + \frac{1}{\mathrm{e}}\ln(c + \mathrm{e}l) - \frac{1}{\mathrm{e}}\ln\left(c + \frac{\mathrm{e}\,l}{2}\right)\right]} \tag{38}$$

The constants and Kapitza conductance values are listed in Table. 2.

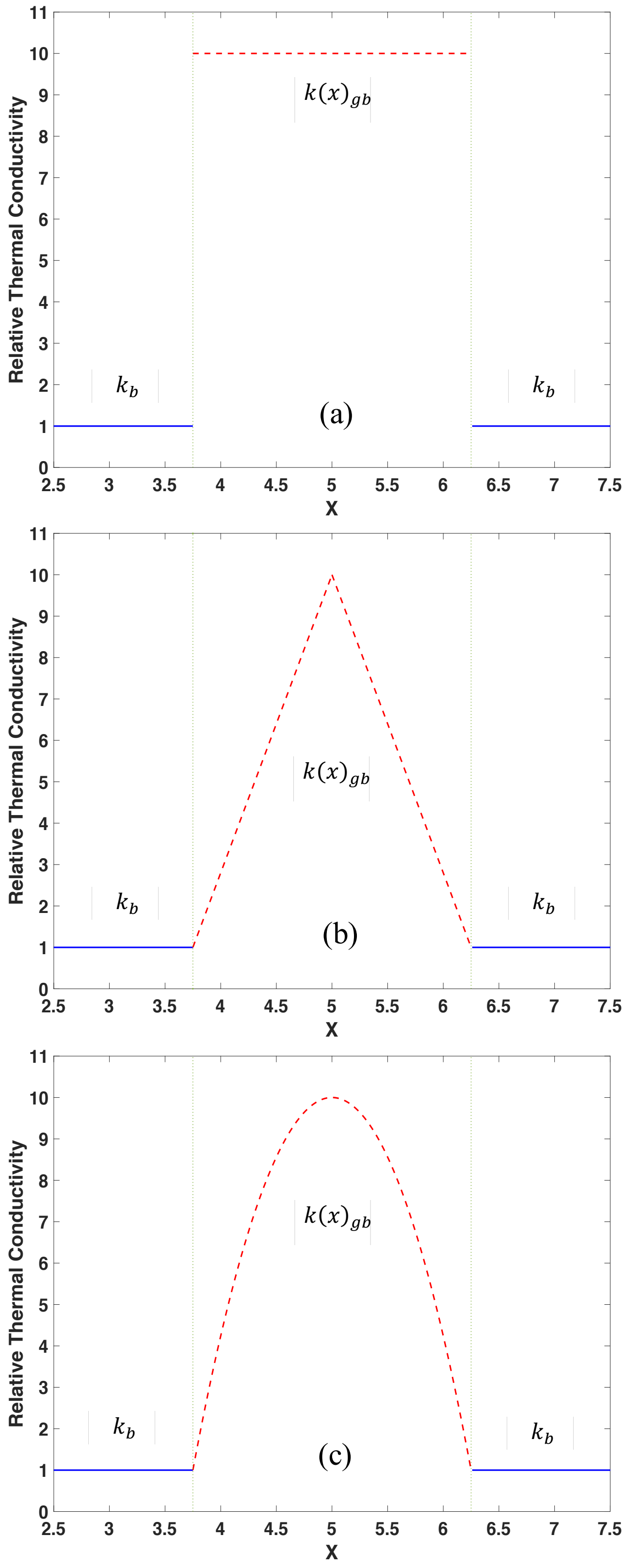

Figure 5. Different spatial profiles for the thermal conductivity of the highly conducting grain boundary: (a) constant, (b) linear, and (c) parabolic.

3.1.2.3 <u>Parabolic profile:</u>

$$k(x) = k_b(a - bx^2) \qquad -\frac{l}{2} \le x \le \frac{l}{2} \tag{39}$$

where $a$ $and$ $b$ are constants that take specific values to set the interface thermal conductivity to be10 times higher than the bulk thermal conductivity value at the middle of the GB (see Fig. 5-c). By combining Eqs (24) and (39) we get

$$G_k = \frac{1}{\int_{-\frac{l}{2}}^{\frac{l}{2}} \frac{dx}{k_b(a - bx^2)}} \tag{40}$$

which after integration results in

$$G_k = \frac{k_b\sqrt{ab}}{\left[\ln\left|\frac{l}{2}\sqrt{\frac{b}{a}} - 1\right| - \ln\left|\frac{l}{2}\sqrt{\frac{b}{a}} + 1\right| - \ln\left|-\frac{l}{2}\sqrt{\frac{b}{a}} - 1\right| + \ln\left|-\frac{l}{2}\sqrt{\frac{b}{a}} + 1\right|\right]} \tag{41}$$

The constants and Kapitza conductance values are shown in Table. 2

Table 2: analytical and numerical Kapitza conductance results for highly-conducting grain boundary where, $l = 2.5\ nm$ , and $k_b = 1\ W/m.K$

| Spatial profile | Constants | | | | Kapitza conductance W / $(m^2.k)$ | | |
|---|---|---|---|---|---|---|---|
| | | | | | New formula | FEM | |
| | a | b | c | e | | Δx = 0.5 nm | Δx = 0.025 nm |
| Constant | N/A | N/A | N/A | N/A | $4.0 \times 10^9$ | $4.0 \times 10^9$ | $4.0 \times 10^9$ |
| Linear | 1 | $18/l$ | -18 | $19/l$ | $1.5634\times 10^9$ | $1.5839\times 10^9$ | $1.5634\times 10^9$ |
| Parabolic | 10 | $36/l^2$ | N/A | N/A | $2.0868\times 10^9$ | $2.1429\times 10^9$ | $2.0868\times 10^9$ |

As evident from Table 2, the FEM numerical results agree well with the newly derived Kapitza conductance formula (Eq. (24)) and are in close agreement for all spatial profiles of the GB thermal conductivity. This agreement improves with a decrease in element size (or equivalently increasing the number of elements inside the GB). Again, it is worthy to note that the spatial profile of the thermal conductivity of the GB affects the Kapitza conductance, with the constant profile leading to the highest value and the linear profile resulting in the lowest value. This agrees with the interface engineering method proposed in[13], where the enhancement of thermal conductance was shown to depend on the spatial profile of the mass-graded interlayers.

## 3.2 Effect of the spatial profile of the GB conductivity on the overall thermal conductivity

In the previous subsection we validated the formulae of the effective Kapitza resistance and conductance of a thin interface. It was shown that these values are affected by how the thermal conductivity of the GB changes with position. Hence, here we study the effect of this change on the overall thermal conductivity. We utilize again the simple configuration of a bicrystal. We run finite-element simulations for the three spatial profiles discussed in the preceding subsection for different grain sizes, with an assumed GB width of 2.5 nm. For both the highly-conducting and lowly-conducting cases, we determine the overall/effective thermal conductivity of the bicrystal using Eq. (26) as discussed in section 2.3.

For the case of highly-conducting GB, Fig. 6 summarizes the general trend. As obvious from the figure, the effective thermal conductivity decreases with the increase of grain size for all spatial profiles of the GB thermal conductivity. This is because in this scenario the GB is the highly conductive 'phase'. This behavior is more pronounced for grain sizes below 100 nm. For instance, as the grain size increases from 10 nm to 100 nm, the effective thermal conductivity decreases by 26.1 %, 22.9%, and 20.5% for the constant, parabolic, and linear spatial profiles, respectively.

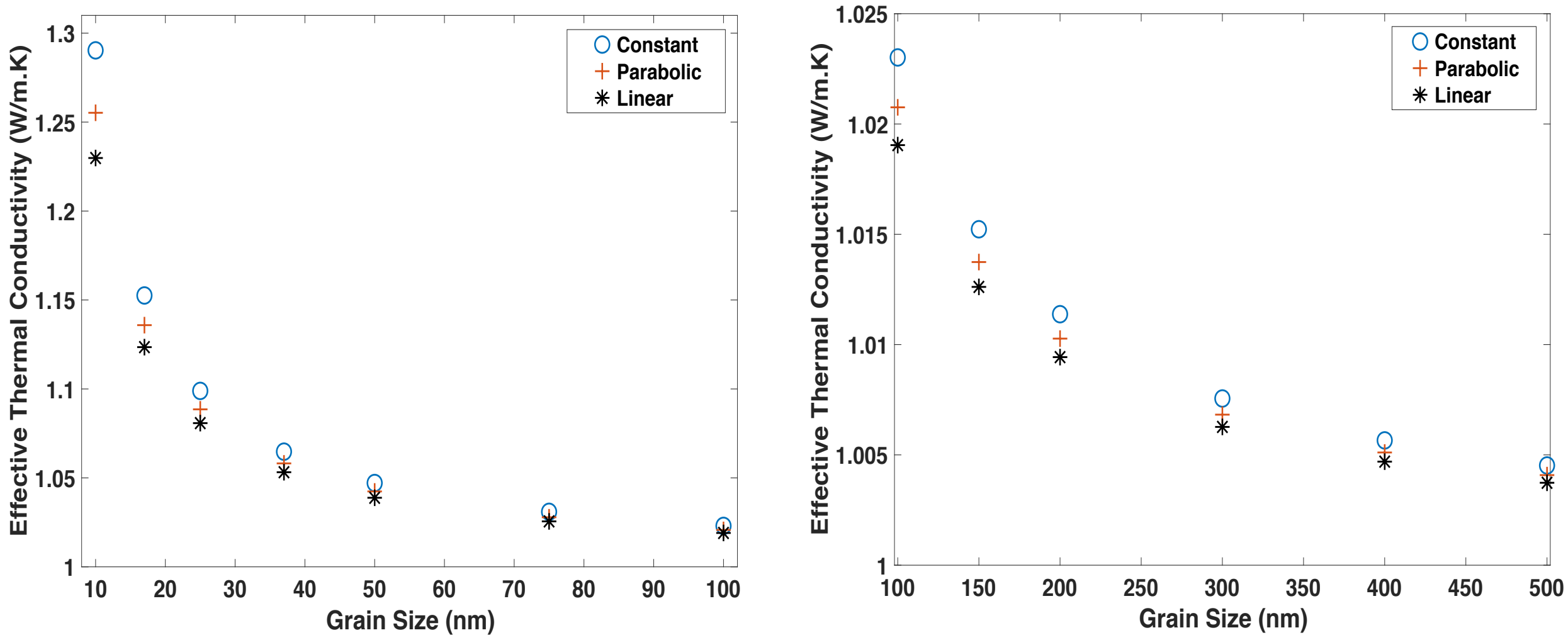

Figure 6. Change of the effective thermal conductivity of a bicrystal with grain size: from 10 nm to 100 nm (left), and from 100 nm to 500 nm (right) for different spatial profiles for the thermal conductivity of the highly-conducting grain boundary.

On the other hand, as the grain size increases from 100 nm to 500 nm, the effective thermal conductivity decreases by 1.84 %, 1.66%, and 1.52% for the constant, parabolic, and linear spatial profiles, respectively. Additionally, Fig. 6 also shows that for the same grain size the spatial profile leads to different values of the effective thermal conductivity, where the highest value is attained by the constant profile and the lowest value is attained by the linear profile. This is because these profiles have the strongest and weakest Kapitza conductance, respectively (see Table. 2). This variation between the results increases as the grain size decreases. For example, the difference between the effective thermal conductivity produced by the constant and linear spatial dependence functions is 5%, 0.4%, and 0.07% at grain size 10 nm, 100 nm, and 500 nm, respectively.

For the case of lowly-conducting GB, the general trend is depicted in Fig. 7.  It shows that, in contrast to the highly-conducting boundary, the effective thermal conductivity increases as the grains size increases for all spatial profiles of grain boundary thermal conductivity. This is due to the fact that the grain boundary acts as an obstacle for heat transport, which is commonly the case. Similar to the case of highly-conducting boundary, this behavior is more apparent in grain sizes below 100 nm. For instance, as the grain size increases from 10 nm to 100 nm, the effective thermal conductivity increases by 164 %, 65%, and 32.5% for the constant, parabolic, and linear spatial profiles of the GB thermal conductivity, respectively. However, as the grain size increases from 100 nm to 500 nm, the effective thermal conductivity increases by 17.6 %, 6.3%, and 3.2% for the constant, parabolic, and linear spatial profiles, respectively.

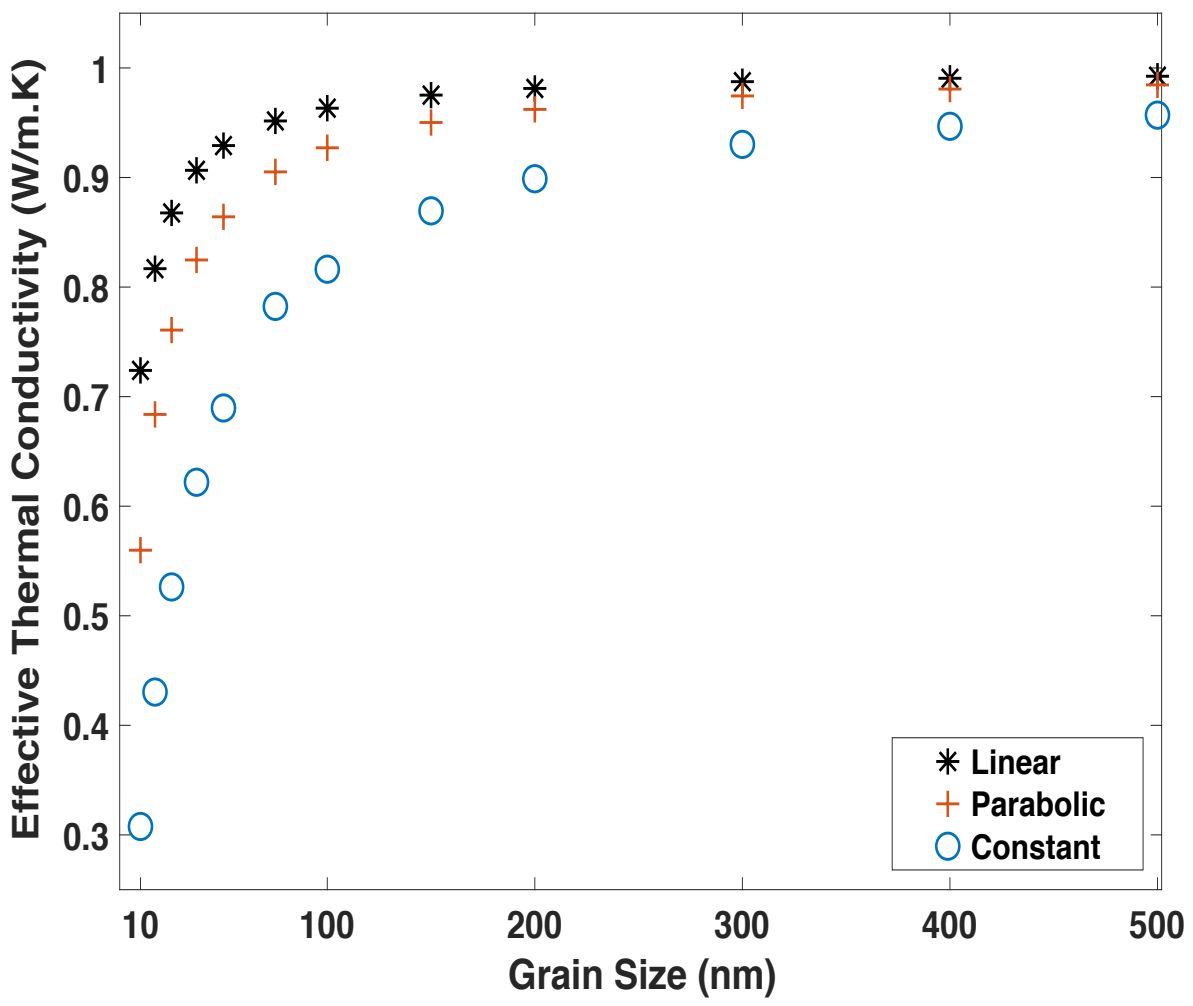

Figure 7. Change of the effective thermal conductivity of a bicrystal with grain size for different spatial profiles of the thermal conductivity of the lowly-conducting grain boundary.

Thus, different from the case of highly-conducting GB, the effective thermal conductivity is still sensitive to variations in the grain size up to 500 nm.

It is worth noting that Fig. 7 also demonstrates that for the same grain size, the distinct spatial profiles lead to different values of the effective thermal conductivity, where the lowest value is attained by the constant profile and the highest value is attained by the linear profile. This is in agreement with the predictions listed in Table 1, where the strongest and weakest Kapitza resistances are obtained for the constant and the linear profiles, respectively. Again, similar to the case of highly-conducting boundary, the variation of the effective conductivity with the spatial profiles decreases as the grain size increases. For example, the difference between the effective thermal conductivity attained by the constant and linear profiles is 135%, 18.2%, and 3.7% at grain size 10 nm, 100 nm, and 500 nm, respectively. Therefore, in general, for all cases, the sensitivity of the effective thermal conductivity to the GB thermal conductivity reduces with increasing grain size due to the decrease of GB area per unit volume.

### 3.3 Comparison of the predictions of the sharp-interface and thin-interface based models

In this subsection, we compare the predictions of the effective thermal conductivity models based on the sharp-interface (Eq. (2)) and the thin-interface (Eq. (15)) descriptions. We utilize again the simple bicrystal configuration and study the effects of grain size, GB width, and Kapitza resistance. Clearly, the effect of GB width can appear only in the thin-interface based model. Hence, first, to investigate this effect, we run finite-element simulations of the bicrystal configuration with different GB width and grain size. In these simulations, the bulk thermal conductivity was 1.0 W/m.K and the GB thermal conductivity was 0.35 W/m.K.

The results of these simulations are shown in Fig. 8. Generally speaking, the effective thermal conductivity decreases as the GB thickness increases regardless of grain size. While the physical width of a typical GB is usually less than a few nanometers, we expand the results for tens of nanometers to account for scenarios as segregation or grain boundary phase formation that may occur. The dependence of the effective thermal conductivity on the GB width is nonlinear as evident from Fig. 8 and predicted from Eq. (15). Since thermal conductivity of the GB is taken as a constant, the Kapitza resistance can be obtained from Eq. (28). Hence, the interfacial resistance is directly proportional to the GB thickness. Clearly, for the same interface width/Kapitza resistance, grains with larger sizes attain higher effective thermal conductivities. Moreover, the sensitivity of the effective thermal conductivity to the grain boundary thickness is inversely proportional to grain size.

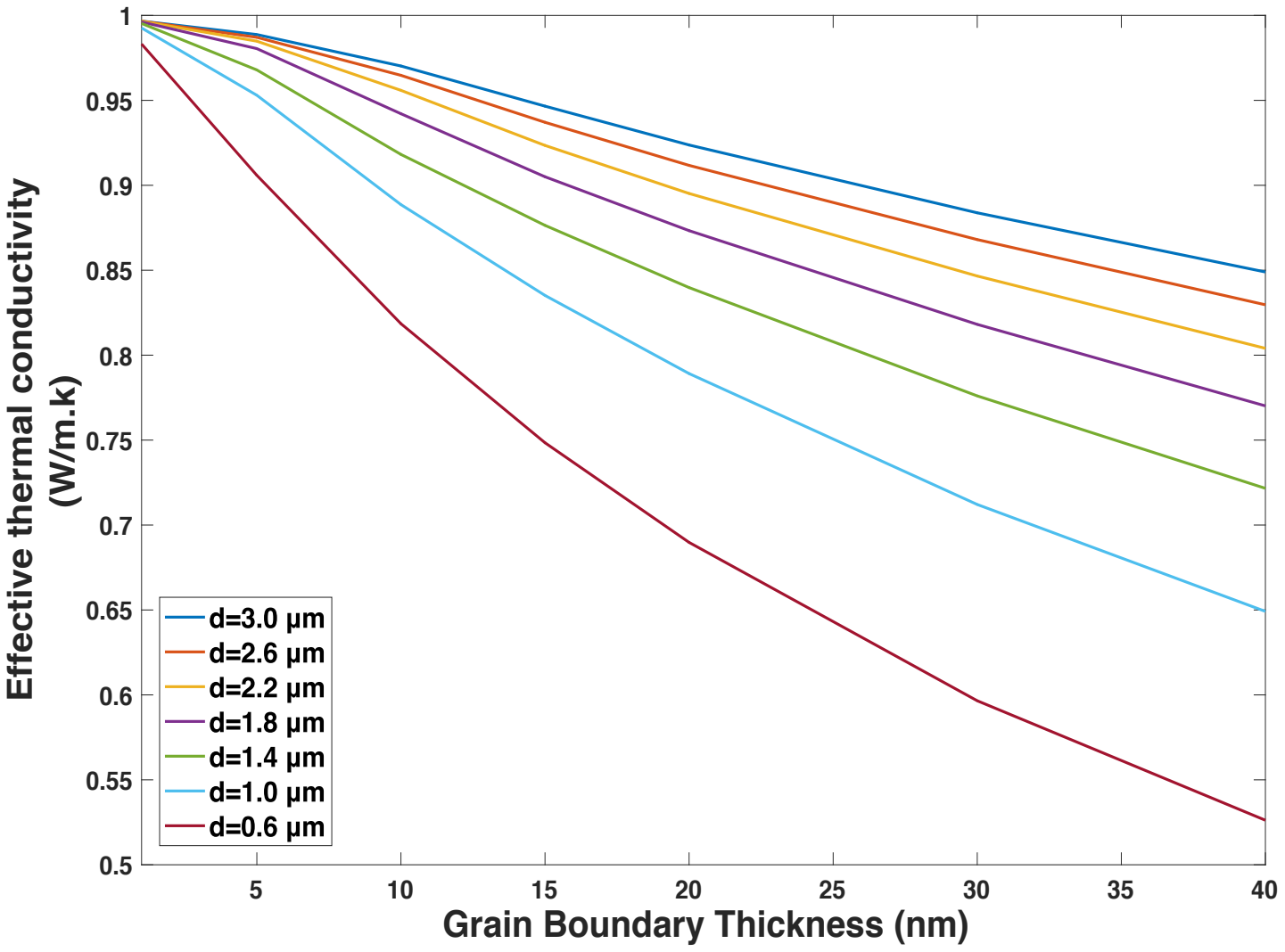

Figure 8. Finite-element simulations of the effect of GB thickness on the effective thermal conductivity of a bicrystal with different grain sizes.

As discussed above, the effects of GB width and thermal resistance are more pronounced for smaller grain sizes. Hence, we now compare the difference in the predictions of the effective thermal conductivity between Yang et al. model (Eq. (2)) and our new model (Eq. (15)) for the case of a nanometer-sized bicrystal. As before, the bulk thermal conductivity was taken to be 1 (W/m.k). The effective conductivity is then calculated based on the above-mentioned equations as function of Kapiza resistance and grain size. In our model, the GB width was set to 2 nm. The results are summarized in Fig. 9. First, the effective thermal conductivity is plotted against the Kapitza resistance, where the grain size was fixed at 20 nm. For both models, the effective thermal conductivity decreases as the Kapitza resistance increases. Moreover, it is clear that the new model predicts higher values of the effective thermal conductivity than Yang et al. model, where the difference in values is directly proportional to the Kapitza resistance. For example, the difference increased from 0% to 10% across the range of Kapitza resistance studied. Second, the effective thermal conductivity is plotted against the grain size in the range 2-20 nm, where the Kapitza resistance was fixed at $20 \times 10^{-9}\, m^2.\frac{k}{W}$. For both models, the effective thermal conductivity increases as the grain size increases. Again, the new model always predicts higher values of the effective thermal conductivity than Yang et al. model for all grain sizes. The difference in the values is inversely proportional to the grain size. For instance, the difference increased from 10% to 100% as the grain size decreased from 20 nm to 2 nm. The reason is that, in contrast to the new model, Yang et al. model neglects the effect of GB thickness, which becomes more apparent as the value of the grain size approaches the value of the GB width.

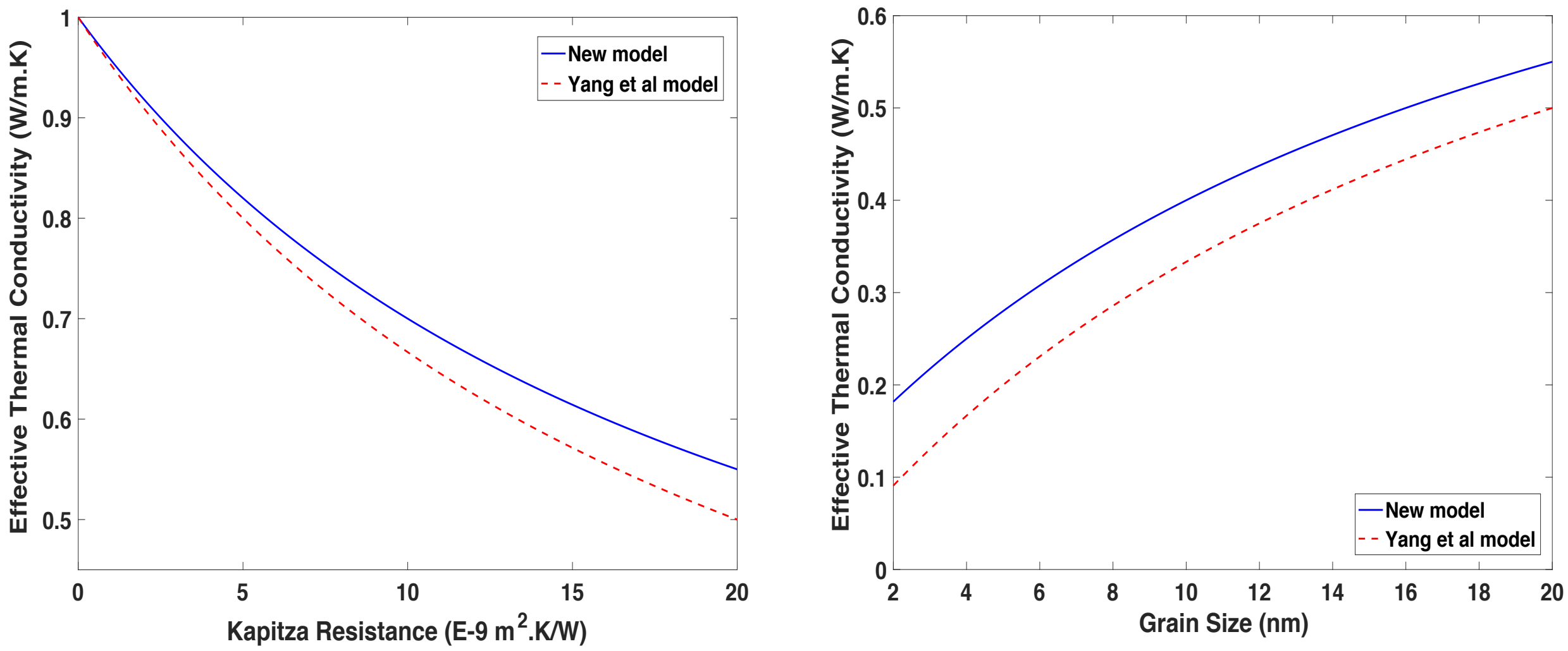

Figure 9. Comparison of the predictions of the new and Yang et al. models for the effective thermal conductivity of a nanometer-sized bicrystal. The effect of the Kapitza resistance (left) and the effect of grain size (right).

The plots in Fig. 9 were combined into 3D figures to better illustrate the behavior of the two models against variations in both grain size and Kapitza resistance concurrently. These 3D plots are presented in Fig. 10. Both models show qualitatively similar trends. The effective thermal conductivity decreases as the Kapitza resistance increases. For the same Kapitza resistance, the effective thermal conductivity increases with grain size. Moreover, the sensitivity of the effective thermal conductivity to grain size variation is directly proportional to the Kapitza resistance. For instance, for the new model, as the grain size increased from 2 nm to 20 nm, the effective thermal conductivity increased by 36.7% and 205.5% for the Kapitza resistance values of $1\times 10^{-9}$ and $20 \times 10^{-9}\, m^2.\frac{k}{W}$, respectively.

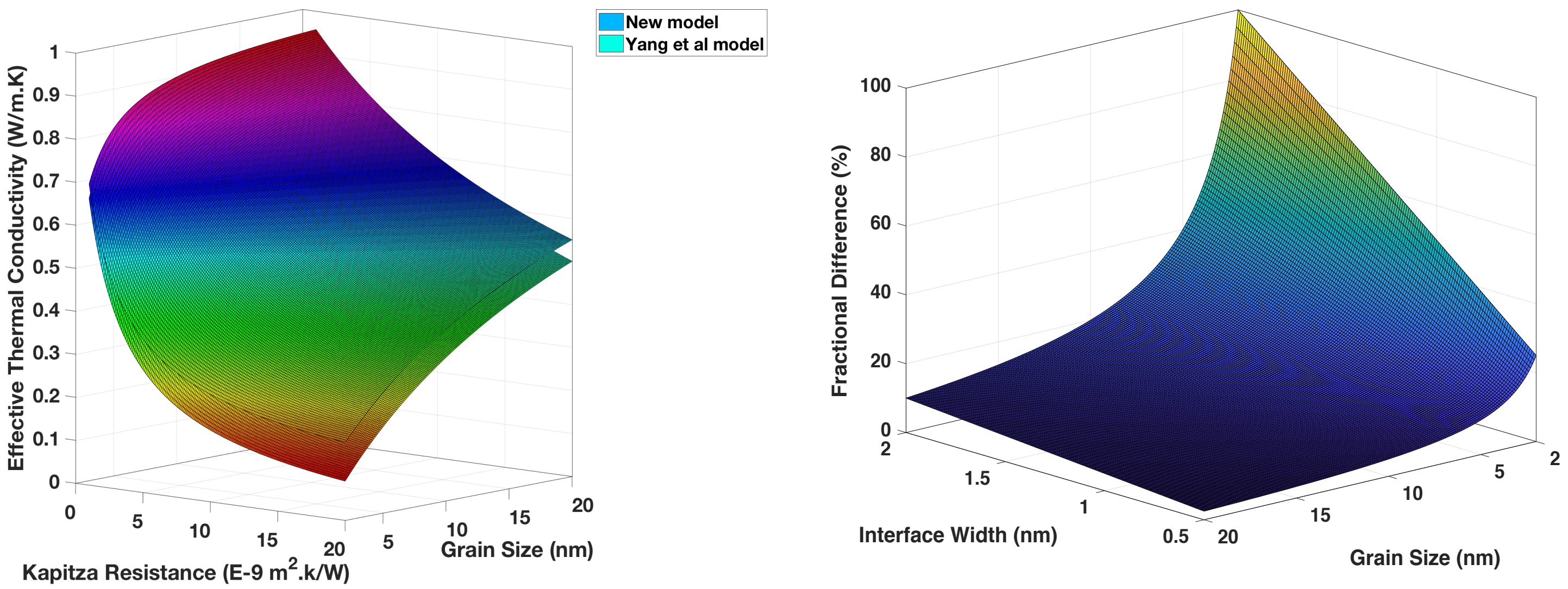

Figure 10. Comparison of the predictions of the new and Yang et al. models for the effective thermal conductivity of a nanometer-sized bicrystal. The effect of the Kapitza resistance and grain size (left) and the effect of GB width (right). The difference increases with increasing Kapitza resistance, increasing GB width, and decreasing grain size.

On the other hand, the sensitivity of effective thermal conductivity to variation in the Kapitza resistance is inversely proportional to grain size. For example, for the new model, as the Kapitza resistance value increased from 0 to $20 \times 10^{-9}\, m^2.\frac{k}{W}$, the effective thermal conductivity was decreased by 74 % and 425.5% for the grain size values of 20 and 2 nm, respectively.

While the two models agree qualitatively, there is a notable quantitative difference in their predictions. The new models always predict higher values of the effective conductivity than Yang et al. model. The deviation between the models increases with decreasing grain size. This is also illustrated in Fig. 10 (right subplot). When the grain size becomes equal to the grain boundary width, a maximum fractional difference of 100% is realized (recall Eq. 18).

### 3.4 Comparison of the predictions of the analytical models based on data from simulations and experiments

To further investigate the differences between the analytical models, we compare their predictions using new finite-element simulations and existing results from molecular dynamics simulations and experiments. Note that similar to Yang et al. model, our model can predict either the effective thermal conductivity given the Kapitza resistance or the Kapitza resistance given the effective thermal conductivity. We will utilize those two indicators in comparing the models.

### 3.4.1 *Comparison of models using FEM simulations of a bicrystal*

We compare here the predictions of the analytical models for the effective thermal conductivity value as function of Kapitza resistance for specific grain sizes. Two different grain sizes of 200 $nm$ and 20 $nm$ were considered. Finite-element simulations of a bicrystal with different values of the Kapitza resistance were performed. The bulk thermal conductivity and GB width were taken as 1 W/m.k and $2nm$, respectively. Fig. 11 summarizes the results of the analytical models and the finite-element simulations. As evident from the figure, the new model is always closer to the results from the numerical simulations than Yang et al. model for all values of grain size and Kapitza resistance. The deviation becomes more apparent for smaller grain sizes. The difference between both analytical models is 1% for the 200 nm grain size and 20% for the 20 nm grain size. Moreover, Fig. 11 also shows that, for the same effective conductivity, the new model always predicts higher values for the Kapitza resistance for all grain sizes.

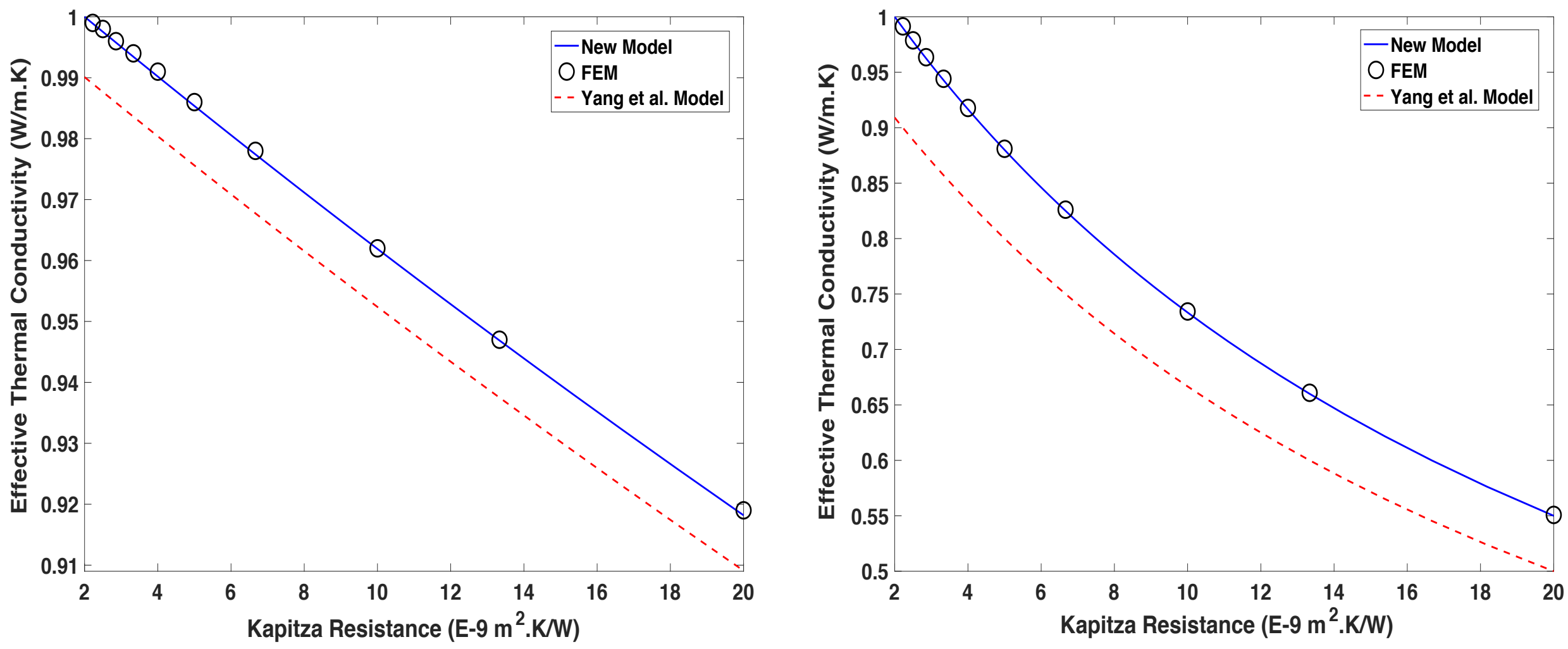

Figure 11. Comparison between the predictions of our new model, Yang et al. model, and FEM simulations for a bicrystal with grain sizes of 200 nm (left) and 20 nm (right).

### *3.4.2 Comparison of models using molecular dynamics simulations of solid argon*

Here, we compare the models with the molecular dynamics simulations (MD) of nanocrystalline argon reported in [28]. To keep argon in solid state, where it assumes an FCC crystal structure, the temperature was set to 30 K [28]. Based on the literature, the GB width is about 1-2 nm for similar materials [57-59]. Hence, the Kapitza resistance will be determined for that range. To achieve that goal, for a specific GB width, we fit our model (Eq. (15)) to the effective thermal conductivity calculations from the MD simulations. An example of this fitting procedure is shown in Fig. 12. The two sharp-interface based models (Eq. (1) and Eq. (2)) were also fitted against the MD results. The predicted values of the Kapitza resistance for the three analytical models are summarized in Table 3. Based on the results listed in Table 3, the Kapitza resistance predicted by our new model is higher than the value from Yang et al. model by 44.3 % for a 1 nm boundary width and by 88% for 2 nm boundary width.

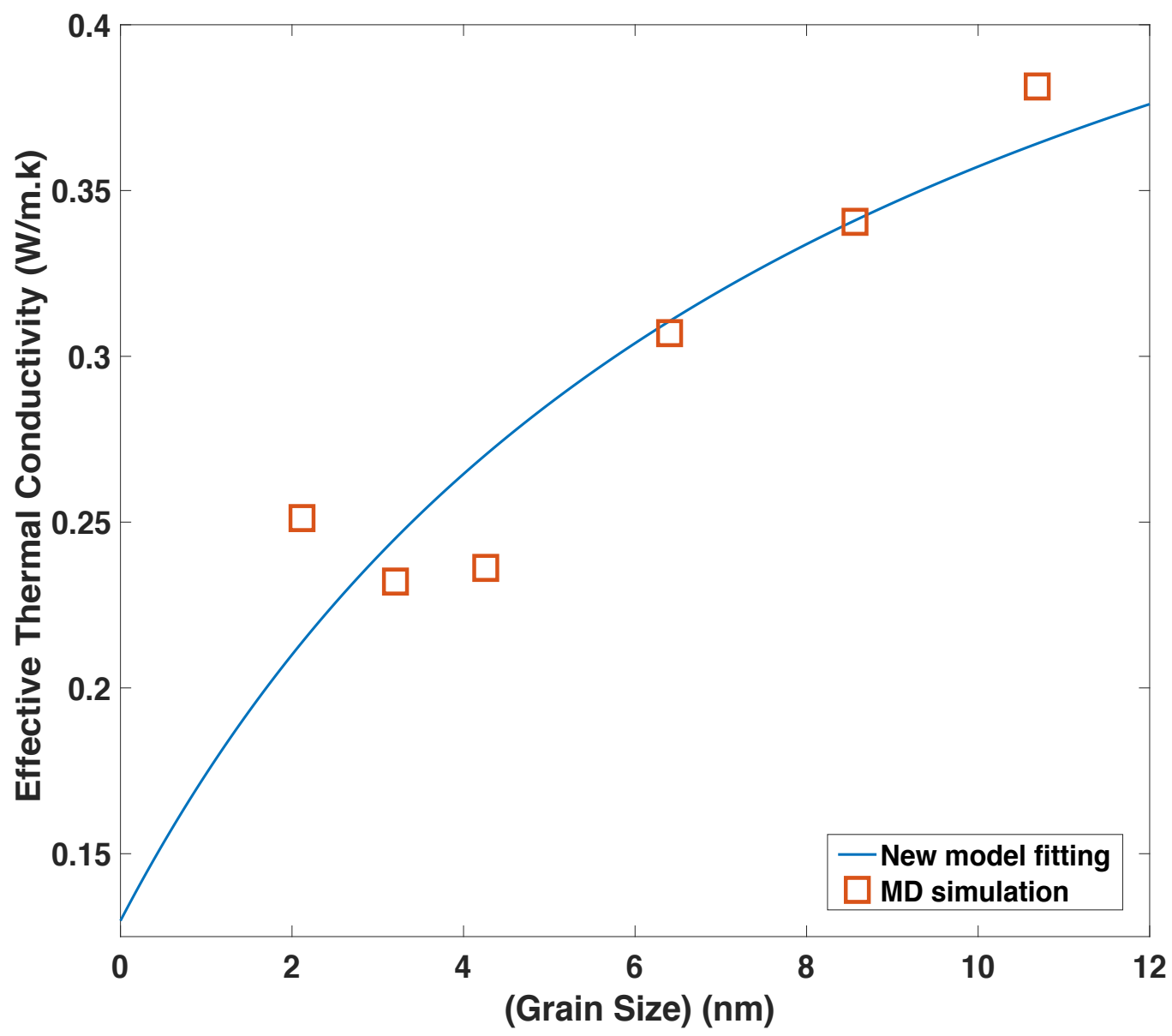

Figure 12. Fitting of the results of molecular dynamics simulations for the effective thermal conductivity of solid argon at 30 K. The solid line is the best fit to the new model (Eq. (15)).

Table 3. The Kapitza resistance predicted by the different analytical models.

| Fitted model | | $l$ $(nm)$ | $R_k$ (m$^2$ K/W) |
|---|---|---|---|
| Nan and Birringer | $k = \frac{k_b}{1 + 2R_k k_b/d}$ | N/A | 4.096$\times 10^{-9}$ |
| Yang et al. | $k = \frac{k_b}{1 + R_k k_b/d}$ | N/A | 8.192 $\times 10^{-9}$ |
| New model | $k = \frac{(d+l)}{\frac{d}{k_b} + R_k}$ | 1 | 11.82 $\times 10^{-9}$ |
| | | 1.5 | 13.62 $\times 10^{-9}$ |
| | | 2 | 15.41 $\times 10^{-9}$ |

### *3.4.3 Comparison of models using experimental data of platinum*

Here, we compare the predictions of the models using experimental data for in-plane thermal conductivity of polycrystalline platinum nanofilms, reported in [60] . The thickness of the nanofilms ranges from 15.0 to 63.0 nm and the mean grain size measured by x-ray diffraction varies between 9.5 and 26.4 nm [60]. To determine the Kapitza resistance, the three analytical models (Eqs. (1), (2), and (15)) were fitted against the experimental data of platinum. An example of this fitting procedure is shown in Fig. 13. The predicted values of the Kapitza resistance for the three analytical models are listed in Table 4. The Kapitza resistance predicted by our new model is higher than the value from Yang et al. model by 10.8% for a 1 nm boundary width and by 21.5% for a 2 nm boundary width.

A final comment on the differences between the analytical models is now in order. As we mentioned before, the first model proposed by Nan and Birringer misses the fact that a GB is shared by two grains [9]. Therefore, that model underestimates the Kapitza resistance by a factor of two as was shown by Yang et al. [8]. Our new model takes another step forward by taking into consideration the effect of the GB width, which makes it more suitable to model nanocrystalline solids.

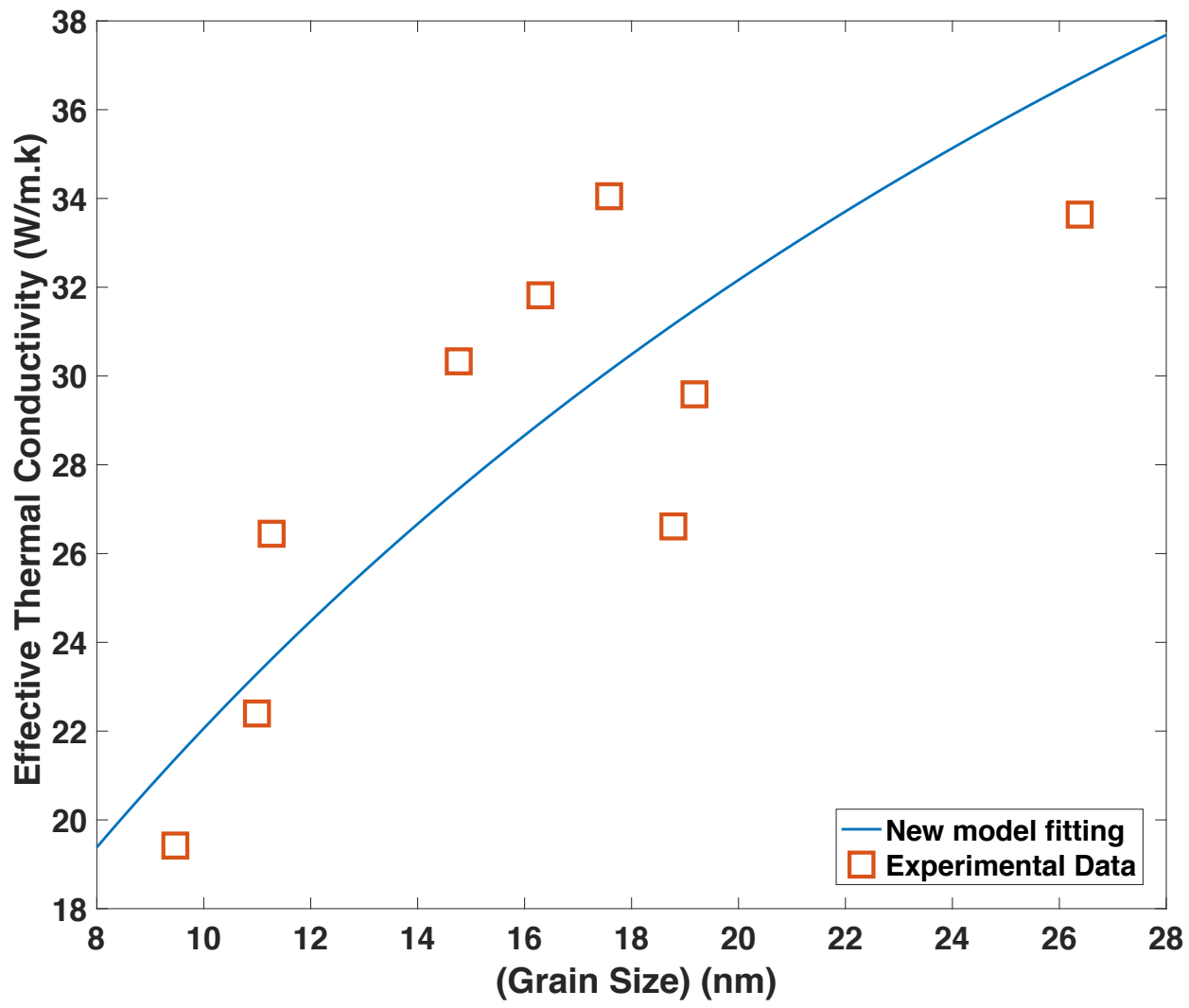

Figure 13. Fitting of the experimental data for the effective thermal conductivity of platinum. The solid line is the best fit to the new model (Eq. (15)).

Table 4. The Kapitza resistance predicted by the different analytical models.

| Fitted model | | $l\ (nm)$ | $R_k$ ($m^2$ K/W) |
|---|---|---|---|
| Nan and Birringer | $k = \dfrac{k_b}{1 + 2R_k k_b / d}$ | N/A | $0.1662 \times 10^{-9}$ |
| Yang et al. | $k = \dfrac{k_b}{1 + R_k k_b / d}$ | N/A | $0.3324 \times 10^{-9}$ |
| New model | $k = \dfrac{(d + l)}{\dfrac{d}{k_b} + R_k}$ | 1 | $0.3682 \times 10^{-9}$ |
| | | 1.5 | $0.3861 \times 10^{-9}$ |
| | | 2 | $0.4039 \times 10^{-9}$ |

## 3.5 Simulations of the effective thermal conductivity of polycrystalline solids

Here we validate our new model for the case of polycrystalline solids using representative microstructures. To that end, we use the phase-field method to generate and evolve these microstructures. Specifically, we utilize the phase-field model of grain growth that is implemented in the open-source MOOSE framework[61]. We investigate two microstructures: (1) a static microstructure with hexagonal grains, (2) a dynamic microstructure with equiaxed grains.

### *3.5.1 Effective thermal conductivity of static hexagonal grains*

We investigate first the effective thermal conductivity of a static microstructure with hexagonal grains. Since hexagonal grains have straight sides, curvature-driven grain growth cannot take place. Thus, steady state calculations were performed using a bulk conductivity of 1 W/mK, GB conductivity of 0.1 W/mK, and a GB width of 2 nm. The results of the finite-element simulations and the analytical models are compared in Fig. 14. As can be seen from the figure, the effective thermal conductivity increases with grain size because of the reduction in the total GB area as the grain size increase. It is also clear from the figure that the predictions of the new model agree more closely with the FEM results than the model of Yang et al. For instance, the difference between the results of Yang et al. model and FEM varied from 2.73 % to 11.48 % across the grain size range studied.  On the other hand, the difference between the results of the new model and FEM varied from 0.03 % to 0.24 % across the same grain size range.

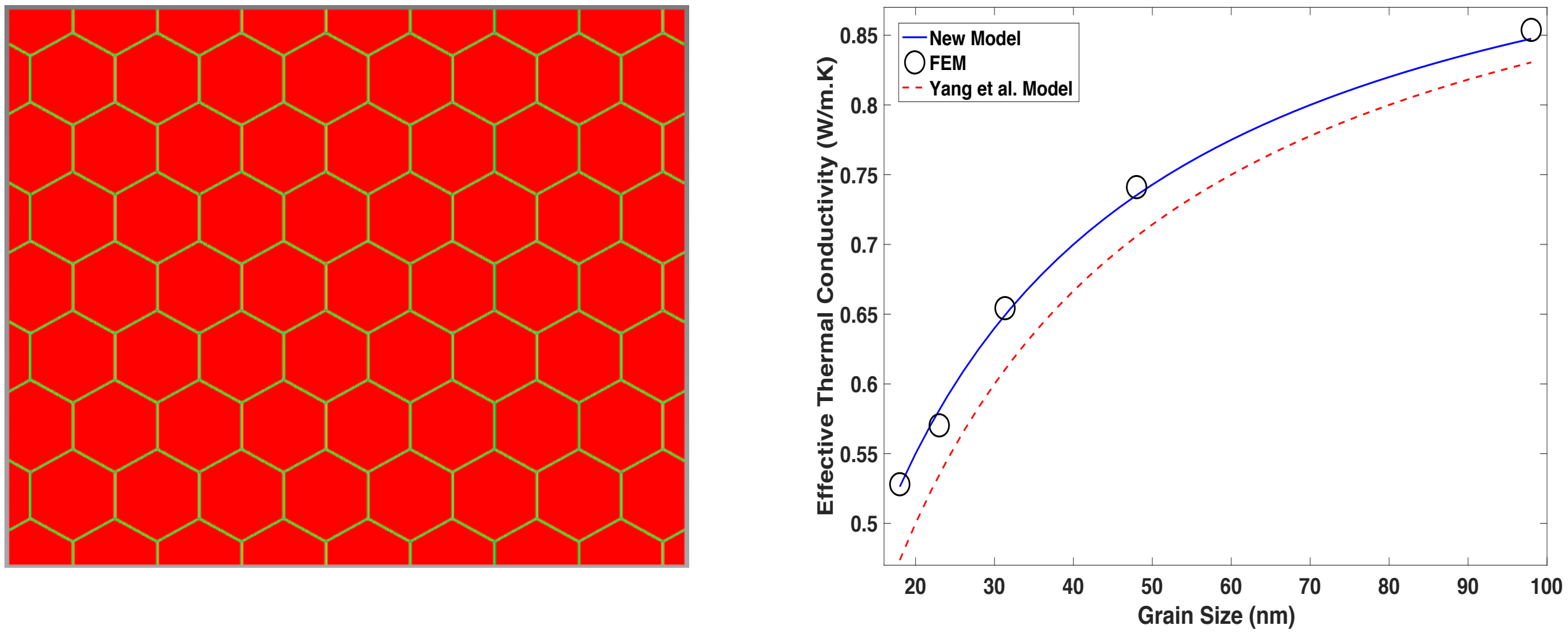

Figure 14. Effective thermal conductivity of a polycrystalline solid with hexagonal grains: (left) a snapshot of the microstructure, (right) the quantitative change of the effective thermal conductivity with grain size.

As we discussed before, another way to evaluate the models is to compare their predictions of the Kapitza resistance instead of the effective conductivity, similar to the work presented in subsection 3.4 . From the FEM simulations, the Kapitza resistance can be calculated directly from the flux and temperature gradient across the grain boundary as given by Eq. (7). For the analytical models, the Kapitza resistance can be obtained by fitting the equations of the models given the grain size and effective conductivity data from the finite-element simulations. The value of the Kapitza resistance obtained directly form the FEM simulation was $20 \times 10^{-9}\, m^2.\frac{k}{W}$ . The predictions of the analytical models are listed in Table 5. As evident from the table, the new model displays better agreement with the numerical value.

Table 5. Errors in the values of the Kapitza resistance obtained from the analytical models

| Fitted Model | | $R_k\ (m^2.\frac{k}{W})$ | Error |
|---|---|---|---|
| Yang et al Model | $k = \frac{k_b}{1 + R_k k_b / d}$ | 16.67$\times 10^{-9}$ | 16.65 % |
| New model | $k = \frac{(d+l)}{\frac{d}{k_b} + R_k}$ | 19.94$\times 10^{-9}$ | 0.3 % |

### 3.5.2 *Effective thermal conductivity of evolving equiaxed grains*

Here, we study the changes of the effective thermal conductivity in polycrystalline solids undergoing grain growth. Voronoi tessellation is utilized to approximate the shapes of random equiaxed grains. Since the effective conductivity is sensitive to the grain size, its value is expected to vary during grain growth. To capture this variation, we solve the steady-state heat conduction equation along with the phase-field kinetic equations of grain growth[62].

The grain growth model parameters are the same as in[62], and the parameters for the steady-state heat conduction calculations are as follows: 1 W/mK for the bulk conductivity, 0.1 W/m.K for the GB conductivity, and 2 nm for the GB width.

During grain growth, to reduce the interfacial energy of the system, the number of grains decreases and the mean grain size increases. As a result, the effective thermal conductivity increases as the process continues. This is captured in Fig. 15 that shows both the microstructure evolution and the change in the thermal conductivity. Again, the results show the effective thermal conductivity predicted by the new model agrees more closely with FEM than Yang et al. model. For instance, the difference between the results of Yang et al. model and FEM varied from 5% to 6 % across the grain size/evolution time range studied. On the other hand, the difference between the results from our new model and FEM varied only from 0.4 % to 1.3 % across the same range studied.

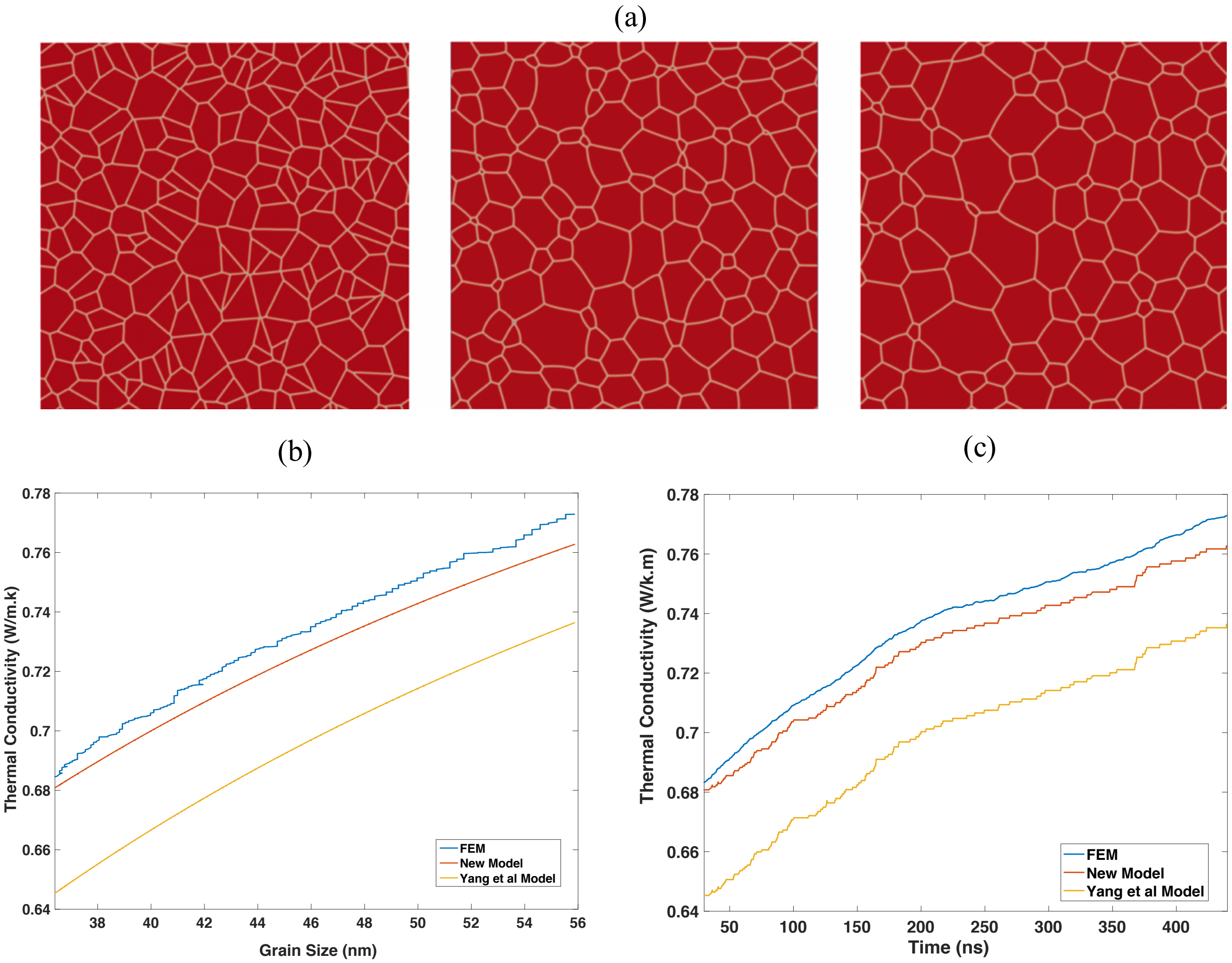

Figure 15. Variation of the thermal conductivity of a polycrystalline solid during grain growth. (a) Snapshots of the microstructure evolution, and the corresponding change of the effective thermal conductivity with (b) grain size and (c) time.

### 3.6 Derivation of a scaling scheme to accelerate mesoscale phase-field simulations

As we have seen in the previous subsection, one can utilize the phase-field method to track both the evolution of microstructure and thermal conductivity of solids. However, running simulations for microstructures with large grain sizes becomes unfeasible if the physical width of GB is used. Nonetheless, based on the derivations we presented above, one can devise an approach to scale up the numerical value of the interface width without changing the effective conductivity. This can be accomplished as follows. Based on the new Kapitza formula (Eq. 23), by assuming constant GB thermal conductivity, the new model (Eq. 15) gives the effective conductivity as,

$$k = \frac{(d+l)}{\frac{d}{k_b} + \frac{l}{k_{gb}}} \tag{42}$$

Therefore, one can first calculate the effective conductivity based on the actual value of the GB width and its corresponding GB conductivity (for a given value of the Kapitza resistance). Then, we utilize the same relation (Eq. (42) to calculate a new scaled value for the GB conductivity that corresponds to the new value of GB width, which we are free to pick for numerical convenience.

As an example for implementing this procedure, We pick the following physical values for the parameters in eq. (42): $d$ = 1.25 $\mu m$ , and $l$ = 1 nm, $k_b and\ k_{gb}$ equal to 4.5 W/m. K, and 0.35 W/m. K, respectively. By substitution of these values into Eq. (42), we obtain $k$ =4.45786 W/m. K. We then perform FEM simulation with the same values of the parameters. From the simulation, the effective thermal conductivity was found to be $k$ =4.457862 W/m. K. Now, the same result can be obtained with a much larger GB width and appropriately scaled GB conductivity. By rearranging Eq. (42), we obtain

$$k_{gb} = \frac{l}{\frac{(d+l)}{k} - \frac{d}{k_b}} \tag{43}$$

Hence, if we choose a GB width of 50 nm instead of 1 nm, Eq. (43) can be used to calculate the corresponding GB conductivity that will result in the same effective conductivity. In this specific example, this value is $k_{gb} = 3.612243\ W/m.k$. To validate this approach, the FEM simulation was repeated again with the scaled values of the GB conductivity and width. The results of the FEM simulations are listed in Table 6. As evident from the table, the scaled values give rise to the same effective conductivity calculated based on the original values, and hence it proves the validity of the scaling approach proposed here.

Table 6: Validation of the scaling approach for enabling large scale simulations.

| model | Original interface width (nm) | Original interface thermal conductivity. $W/(m.K)$ | Scaled interface width (nm) | Scaled interface thermal conductivity. $W/(m.K)$ | Effective Thermal conductivity. $W/(m.K)$ |
|---|---|---|---|---|---|
| FEM | 1 | 0.35 | 50 | 3.612243 | 4.457862 |

## 4 Concluding remarks

A novel model for the effective thermal conductivity of polycrystalline solids was developed. In contrast to existing models, our model is based on the thin-interface description of boundaries. This treatment leads to two major advantages over the classical sharp-interface description. First, it enables the model to predict the thermal conductivity of nanocrystalline materials, where the grain size is comparable to the GB width. Second, it allows the model to simulate the enhancement or degradation of interfacial transport due to segregation, interface roughness, interface strengthening, or interface phase transition. In our derivation of the model, we introduced a general expression for the effective Kapitza resistance/conductance of a thin interface. This new expression was validated using finite-element simulations for different GB thermal conductivity profiles. This new continuum-based treatment of interfacial heat transport is expected to contribute to the understanding of optimizing heat transport via the different methods of interface engineering, which are usually only modeled using atomistic simulations [12, 13, 35, 37].

The predictions of the new model were compared with the existing analytical models. It was shown that the thin-interface based model predicts higher values for the effective thermal conductivity and Kapitza resistance than its sharp-interface counterparts. These predictions were verified using finite-element simulations. For nanograins, the new model predicts 10%-100% higher values of the effective thermal conductivity than Yang et al. model [8], as the grain size approaches the width of the grain boundary. Moreover, the new model was shown to be capable of describing the change of thermal conductivity with microstructure. By coupling the heat-conduction and phase-field equations, finite-element simulations of the co-evolution of microstructure and thermal conductivity were performed for different grain structures. The predictions of the new models were closer to the results from the FEM simulations than those given by the model of Yang et al. [8]. While, we focused here on heat transport across grain boundaries in polycrystalline materials, we also discussed in the appendix how this new approach can be generalized to the case of hetero-interfaces in multiphase materials. However, a complete treatment of this case is beyond the scope of this study and will be investigated in the future.

Lastly, as with any model, this new model has some limitations. Two main limitations can be recognized. First, the model ignores quantum effects that might alter the transmission and scattering probabilities of heat carries in the vicinity of interfaces. However, it is well known that these effects are of importance only in the low temperature regime. Second, the model assumes the validity of Fick's law to describe heat conduction. While this is difficult to justify for nanocrystalline materials, several non-equilibrium molecular dynamics simulations demonstrated that it is a reasonable assumption[12, 13, 35, 37]. This assumption is also employed in most of existing analytical/continuum models and even experimental studies of heat transport in heterogenous solids[1, 8-11]. Nonetheless, our main approach can be adapted to other constitutive laws of interest. For instance, one can utilize the concepts of extended irreversible thermodynamics to derive more appropriate constitutive laws that account for non-local, non-linear, and ballistic effects on interfacial heat transport[63].

**Acknowledgement**

We acknowledge the funding from the Department of Energy Office of Nuclear Energy (DE-NE0008764).

**Data Availability Statement**

The data that support the findings of this study are available from the corresponding author upon reasonable request.

## Appendix A

We generalize here the analysis given above for the thermal resistance of grain boundaries to hetero-interfaces between different phases. A schematic of this configuration is presented in Fig. A1. We start again by assuming the validity of Fourier's law of conduction,

$$q = -k\frac{dT}{dx} \tag{A1}$$

As shown in Figure. A1, the total temperature drop across the two-phase domain is

$$T_{tot} = T_1 + T_{1-2} + T_2 \tag{A2}$$

By substituting Eq. (A2) into Eq. (A1) and rearranging

$$k = \frac{-q(d_1 + d_{1-2} + d_2)}{T_1 + T_{1-2} + T_2} \tag{A3}$$

Where

$d_1$ : First phase region thickness

$d_{1-2}$ : Hetero-interface thickness

$d_2$ : Second phase region thickness

$T_1$ : Temperature drop through the first phase region

$T_{1-2}$ : Temperature drop across the interface

$T_2$ : Temperature drop through the second phase region

Where $\frac{T_1+T_{1-2}+T_2}{(d_1+d_{1-2}+d_2)}$ is equivalent to $\frac{dT}{dx}$.

For the case of steady-state heat conduction and by the assumption of the continuity of heat flux, Fourier's law holds in each region. Hence, we have the following relations.

For the first phase region,

$$T_1 = \frac{-qd_1}{k_1} \tag{A4}$$

For the interfacial region (the Kapitza interfacial boundary condition),

$$T_{1-2} = -qR_k \tag{A5}$$

For the second phase region

$$T_2 = \frac{-qd_2}{k_2} \quad \text{(A6)}$$

By combining Eqs. (A3-A6), we obtain

$$k = \frac{(d_1 + d_{1-2} + d_2)}{\frac{d_1}{k_1} + R_k + \frac{d_2}{k_2}} \quad \text{(A7)}$$

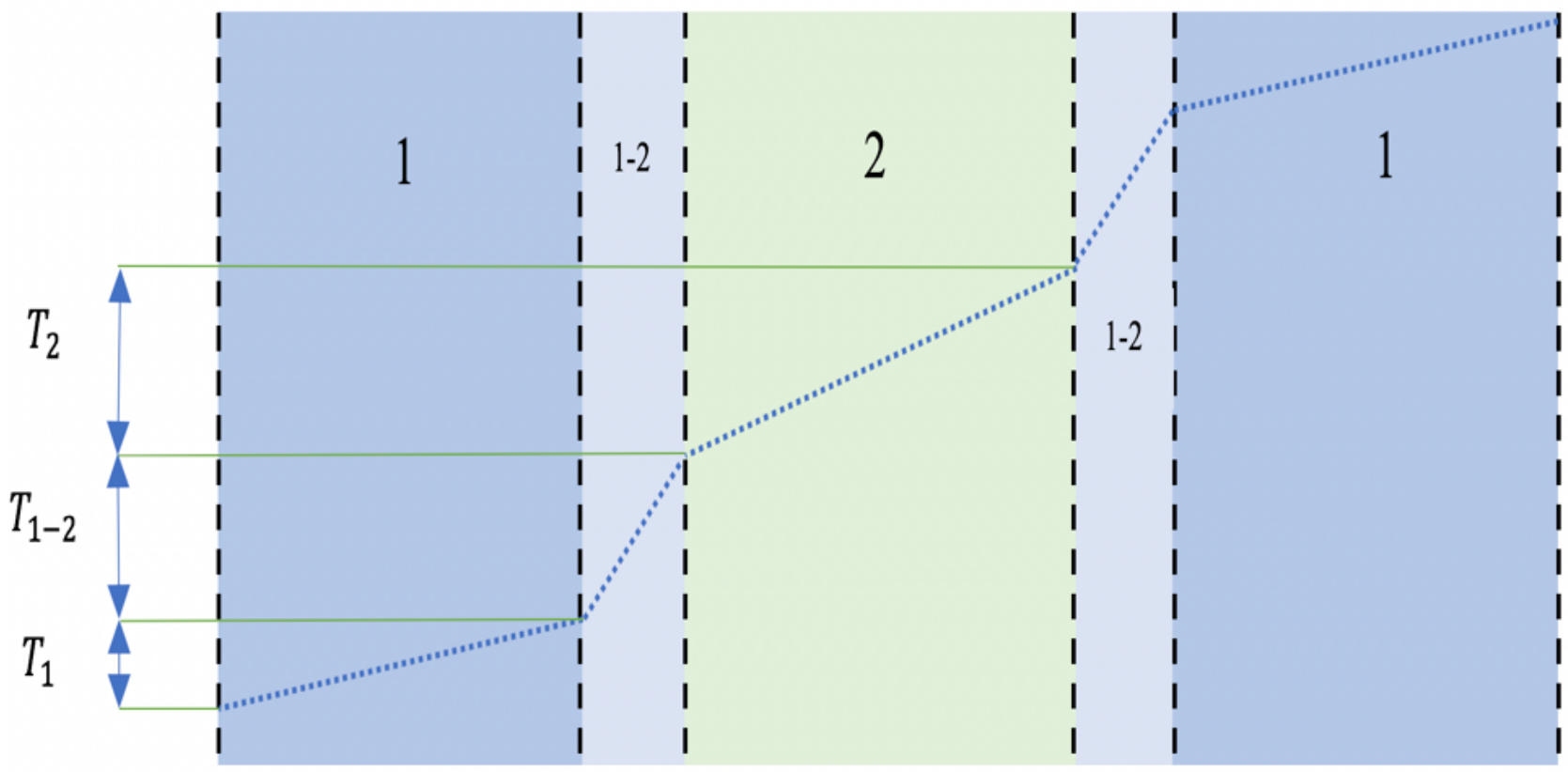

Figure A1. A Schematic representation of a hetero-interface in two-phase systems as a tri-layer composite.

Again, similar to the analysis presented earlier, the same model can be obtained from the harmonic average formula for layered composites (e.g., thermal resistances in series). This is accomplished by simply considering the interfacial region as an autonomous phase. One then has,

$$k = \frac{1}{\frac{\emptyset_1}{k_1} + \frac{\emptyset_{1-2}}{k_{1-2}} + \frac{\emptyset_2}{k_2}} \quad \text{(A8)}$$

Where:

$\emptyset_1$ , $\emptyset_2$ and $\emptyset_{1-2}$ represent the volume fractions of the first phase, second phase, and interface, respectively. Those are defined as follows:

$$\emptyset_1 = d_1/(d_1 + d_{1-2} + d_2) \quad \text{(A9)}$$

$$\emptyset_{1-2} = d_{1-2}/(d_1 + d_{1-2} + d_2) \tag{A10}$$

$$\emptyset_2 = d_2/(d_1 + d_{1-2} + d_2) \tag{A11}$$

By combining Eqs. (A8-A11), we arrive at

$$k = \frac{(d_1 + d_{1-2} + d_2)}{\frac{d_1}{k_1} + \frac{d_{1-2}}{k_{1-2}} + \frac{d_2}{k_2}} \tag{A12}$$

Again, similar to our analysis for the case of grain boundaries, Eq. (A7) and Eq. (A12) have a very similar structure. They are identical if the following relation holds,

$$R_k = \frac{d_{1-2}}{k_{1-2}} \tag{A13}$$

Now recall that we derived a general expression for the effective Kapitza resistance of a thin interface in section 2. The application of that definition for the current case leads to

$$R_k = \int_0^{d_{1-2}} \frac{dx}{k(x)_{1-2}}, \tag{A14}$$

which, for the case of constant thermal conductivity in the interfacial region, reduces to Eq. (A13). One can of course generalize this analysis to the case of multi-phase systems with different hetero-interfaces.

Clearly, the above analysis was derived based on a specific microstructure, e.g., alternating layers of phases. This simplified the analysis significantly since only 1D treatment is required. Nonetheless, similar to the case of the polycrystalline materials treated earlier, the current analysis should apply equally well for configurations with continuous distribution of second phase in isotropic materials. However, a more comprehensive analysis is necessary for the case of dispersed microstructures (with discontinuous distribution of second phase particles). In such case, the second phase particle fraction, distribution, size, and morphology will play significant roles, and a full 3D treatment is needed. This will be the subject of an upcoming publication.